\documentclass[twocolumn]{emulateapj}
\usepackage[]{units}
\usepackage{datetime}
\usepackage{graphicx}
\usepackage{amssymb}
\usepackage{amsmath}
\usepackage{threeparttable}
\usepackage{url}

\newcommand{\spi}{{\it Spitzer}}
\newcommand{\her}{{\it Herschel}}
\mathchardef\mhyphen="2D
\slugcomment{To be submitted to MNRAS}
\shorttitle{Star formation in quasar hosts}
\shortauthors{Zakamska et al.}

\begin{document}

\title{Star formation in quasar hosts \\
and the origin of radio emission in radio-quiet quasars}

\author{Nadia L. Zakamska\altaffilmark{1}, Kelly Lampayan\altaffilmark{1}, Andreea Petric\altaffilmark{2}, Daniel Dicken\altaffilmark{3}, Jenny E. Greene\altaffilmark{4}, Timothy M. Heckman\altaffilmark{1}, Ryan C. Hickox\altaffilmark{5}, Luis C. Ho\altaffilmark{6,7}, Julian H. Krolik\altaffilmark{1}, Nicole P.H. Nesvadba\altaffilmark{8}, Michael A. Strauss\altaffilmark{4}, James E. Geach\altaffilmark{9}, Masamune Oguri\altaffilmark{10,11,12}, Iskra V. Strateva\altaffilmark{13}}
\altaffiltext{1}{Department of Physics \& Astronomy, Johns Hopkins University, Bloomberg Center, 3400 N. Charles St., Baltimore, MD 21218, USA}
\altaffiltext{2}{Gemini Observatories, 670 N. A'Ohoku Pl., Hilo, HI 96720, USA}
\altaffiltext{3}{Laboratoire AIM-Paris-Saclay, CEA/DSM/Irfu, Orme des Merisiers, Bat 709, F-91191 Gif sur Yvette, France}
\altaffiltext{4}{Department of Astrophysical Sciences, Princeton University, Princeton, NJ 08544, USA}
\altaffiltext{5}{Dartmouth College, Dept. of Physics and Astronomy, 6127 Wilder Laboratory, Hanover, NH 03755, USA}
\altaffiltext{6}{Kavli Institute for Astronomy and Astrophysics, Peking University, Beijing 100871, China}
\altaffiltext{7}{Department of Astronomy, School of Physics, Peking University, Beijing 100871, China}
\altaffiltext{8}{Institut d'Astrophysique Spatiale, CNRS, Universit\'e Paris-Sud, Bat. 120-121, F-91405 Orsay, France}
\altaffiltext{9}{Centre for Astrophysics Research, Science \& Technology Research Institute, University of Hertfordshire, Hatfield AL10 9AB, UK}
\altaffiltext{10}{Research Center for the Early Universe, University of Tokyo, 7-3-1 Hongo, Bunkyo-ku, Tokyo 113-0033, Japan}
\altaffiltext{11}{Department of Physics, University of Tokyo, 7-3-1 Hongo, Bunkyo-ku, Tokyo 113-0033, Japan}
\altaffiltext{12}{Kavli Institute for the Physics and Mathematics of the Universe (Kavli IPMU, WPI), University of Tokyo, 5-1-5 Kashiwanoha, Kashiwa, Chiba 277-8583, Japan}
\altaffiltext{13}{Gymnasium Kirschgarten, Hermann Kinkelin-Strasse 10, 4051 Basel, Switzerland}

\begin{abstract}
Radio emission from radio-quiet quasars may be due to star formation in the quasar host galaxy, to a jet launched by the supermassive black hole, or to relativistic particles accelerated in a wide-angle radiatively-driven outflow. In this paper we examine whether radio emission from radio-quiet quasars is a byproduct of star formation in their hosts. To this end we use infrared spectroscopy and photometry from \spi\ and \her\ to estimate or place upper limits on star formation rates in hosts of $\sim 300$ obscured and unobscured quasars at $z<1$. We find that low-ionization forbidden emission lines such as [NeII] and [NeIII] are likely dominated by quasar ionization and do not provide reliable star formation diagnostics in quasar hosts, while PAH emission features may be suppressed due to the destruction of PAH molecules by the quasar radiation field. While the bolometric luminosities of our sources are dominated by the quasars, the 160\micron\ fluxes are likely dominated by star formation, but they too should be used with caution. We estimate median star formation rates to be $6-29$ $M_{\odot}$ yr$^{-1}$, with obscured quasars at the high end of this range. This star formation rate is insufficient to explain the observed radio emission from quasars by an order of magnitude, with $\log(L_{\rm radio,obs}/L_{\rm radio,SF})=0.6-1.3$ depending on quasar type and star formation estimator. Although radio-quiet quasars in our sample lie close to the 8-1000\micron\ infrared / radio correlation characteristic of the star-forming galaxies, both their infrared emission and their radio emission are dominated by the quasar activity, not by the host galaxy. 
\end{abstract}


\section{Introduction}
\label{sec:intro}

The most extended, powerful and beautiful sources in the radio sky are due to synchrotron emission from relativistic jets launched by supermassive black holes in centers of galaxies \citep{urry95}, but only a minority of active black holes produce these structures. At a given optical luminosity of the active nucleus, radio power spans many orders of magnitude, and the exact distribution of radio luminosities remains a matter of continued debate. A particularly intriguing point is whether this distribution is bimodal \citep{ivez02, whit07, kimb11}: does the brighter ``radio-loud'' population show a well-defined luminosity separation from the fainter ``radio-quiet'' group, or is the distribution of radio luminosities continuous (e.g., \citealt{bonc13})? This question goes to the heart of fundamental issues in black hole physics: are weak radio sources associated with supermassive black holes due to relativistic jets which are scaled down from their extended powerful analogs, or are there additional mechanisms for producing radio emission? Are all black holes actually capable of launching a relativistic jet, and do all black holes undergo such a phase?

At $\ga 1''$ resolution, the majority of quasars ($L_{\rm bol}\ga 10^{45}$ erg s$^{-1}$) are point-like radio sources with luminosities $\nu L_{\nu}[1.4{\rm GHz}]\la 10^{41}$ erg s$^{-1}$, and the origin of this emission has been the subject of recent debate \citep{laor08, cond13, huse13, mull13}. The recent finding of a strong proportionality between the radio luminosity of radio-quiet quasars and the square of the line-of-sight velocity dispersion of the narrow-line gas \citep{spoo09, mull13, zaka14} is an exciting development in this topic, offering possible clues as to the nature of the radio emission. These velocity dispersions can reach values that are much higher than those that can be confined by a typical galaxy potential, suggesting that the ionized gas is neither in static equilibrium nor in galaxy rotation. Blue-shifted asymmetries suggest that the gas is outflowing \citep{zaka14}, and interpreting the line-of-sight velocity distribution as due to the range of velocities in the outflow suggests $v_{\rm out}\sim 1000$ km s$^{-1}$. 

The observed correlation between narrow line kinematics and radio luminosity suggests a physical connection between the processes that produce them. One possibility is that compact jets inject energy into the gas and launch the outflows \citep{veil91c, spoo09, mull13}; another is that the winds are driven radiatively, then induce shocks in the host galaxy and the shocks in turn accelerate relativistic particles \citep{stoc92, wang08a, jian10, ishi11, fauc12b, zubo12, zaka14}. 

A completely different approach is followed by \citet{kimb11} and \citet{cond13} who argue that the radio emission in radio-quiet quasars is mostly or entirely due to star formation in their host galaxies. Three arguments could be put forward to support this hypothesis: (i) If the radio luminosity function is bimodal, then something other than scaled-down jets is probably responsible for the radio-quiet sources. (ii) Active galaxies with $L_{\rm bol}\la 10^{45}$ erg s$^{-1}$ tend to lie on the extension of the classical 8-1000\micron\ / radio correlation of the star-forming galaxies \citep{mori10, rosa13}. (iii) The amount of radio emission seen in high-redshift radio-quiet quasars can be explained by star formation rates $20-500 M_{\odot}$ yr$^{-1}$, which (although quite high) seem plausible for the epoch of peak galaxy formation. 

Several arguments can be put forward against this hypothesis: (i) In quasars, the scatter around the radio / infrared relationship is higher than that seen in star-forming galaxies \citep{mori10}. (ii) In quasars the infrared emission can be dominated by the quasar, rather than by the star formation \citep{hony11, sun14}. (iii) The amount of star formation required to explain the observed radio emission in quasars may be higher than that deduced using other methods \citep{lal10, zaka14}. 

\citet{rosa13} demonstrate that in radio-quiet low-luminosity active galactic nuclei (AGN) much of the observed radio luminosity is consistent with star formation in the AGN hosts. The objects in their sample have infrared luminosities $\nu L_{\nu}$[12\micron]$\la 10^{44}$ erg s$^{-1}$. In this paper we examine AGNs with $\nu L_{\nu}$[12\micron] from $\sim 2 \times 10^{43}$ to $\sim 10^{46}$ erg s$^{-1}$, thereby extending the analysis of \citet{rosa13} to luminosities higher by up to two orders of magnitude. Our goal is to determine whether the radio emission of quasars ($\nu L_{\nu}$[12\micron]$\ga 10^{44}$ erg s$^{-1}$, or $L_{\rm bol}\ga 10^{45}$ erg s$^{-1}$ as per bolometric corrections by \citealt{rich06}) is due to the star formation in their host galaxies. 

To this end, we estimate the rates of star formation in the hosts of quasars of different types using \spi\ and \her\ data, and compare the amount of radio emission seen from these objects with that expected from star formation alone \citep{helo85, bell03}. In Section \ref{sec:data} we describe sample selection, datasets and measurements. In Section \ref{sec:photo_analysis}, we use far-infrared photometry to calculate star formation rates, predict the associated radio emission and compare with observations. In Section \ref{sec:spec_analysis} we use mid-infrared spectroscopy for a similar analysis. We discuss various difficulties in measuring star formation rates of quasar hosts in Section \ref{sec:discussion} and summarize in Section \ref{sec:conclusions}. We use a $h$=0.7, $\Omega_m$=0.3, $\Omega_{\Lambda}$=0.7 cosmology.

Throughout the paper, we make a key distinction between far-infrared ($\ga 100\micron$) vs radio correlation and total infrared (conventionally defined over 8-1000\micron\ range) vs radio correlation. For star forming galaxies which show similar infrared spectral energy distributions, these concepts can be used interchangeably, since an accurate estimate of the total infrared luminosity can be obtained from far-infrared fluxes alone (e.g., \citealt{syme08}). However, as we add quasar contribution to both infrared and radio emission, some or all of these relationships might break down, and in particular because of the wide range of quasar spectral energy distributions their far-infrared emission and their total infrared emission are no longer strongly correlated. In Section \ref{sec:conclusions}, we investigate the fate of far-infrared vs radio and total infrared vs radio correlations in the presence of a quasar. 

\section{Samples, observations, data reduction and measurements}
\label{sec:data}

\subsection{Type 2 and type 1 samples}

Our goal is to assemble a large sample of quasars (whether optically obscured or unobscured) for which the host star formation rates can be usefully constrained with existing archival data. Furthermore, because of the sensitivity of the existing radio surveys, in order to probe the radio-quiet population we are restricted to low-redshift quasars, $z<1$. As a result, this work primarily focuses on the analysis of two quasar samples. 

Our first sample consists of \spi\ and \her\ follow-up of obscured (type 2) quasars from \citet{reye08} at $z\la 0.8$. These objects are selected to have only narrow emission lines with line ratios characteristic of ionization by a hidden AGN \citep{zaka03} and are required to have $L_{\rm [OIII]}\ga 10^{41.5}$ erg s$^{-1}$. Of the 887 objects in \citet{reye08} catalog, WISE-3 matches are available for 94\% of the objects and WISE-4 matches for 87\% (some of the remaining 13\% are detected, but cannot be deblended from the nearby contaminants in the WISE-4 band). We calculate 12\micron\ luminosities from the WISE-3 matches, k-correcting using WISE-4 flux if available or using a median WISE-4/WISE-3 index if not \citep{zaka14}. 

\begin{figure}
\includegraphics[scale=0.8, clip=true, trim=0cm 0cm 11cm 11cm]{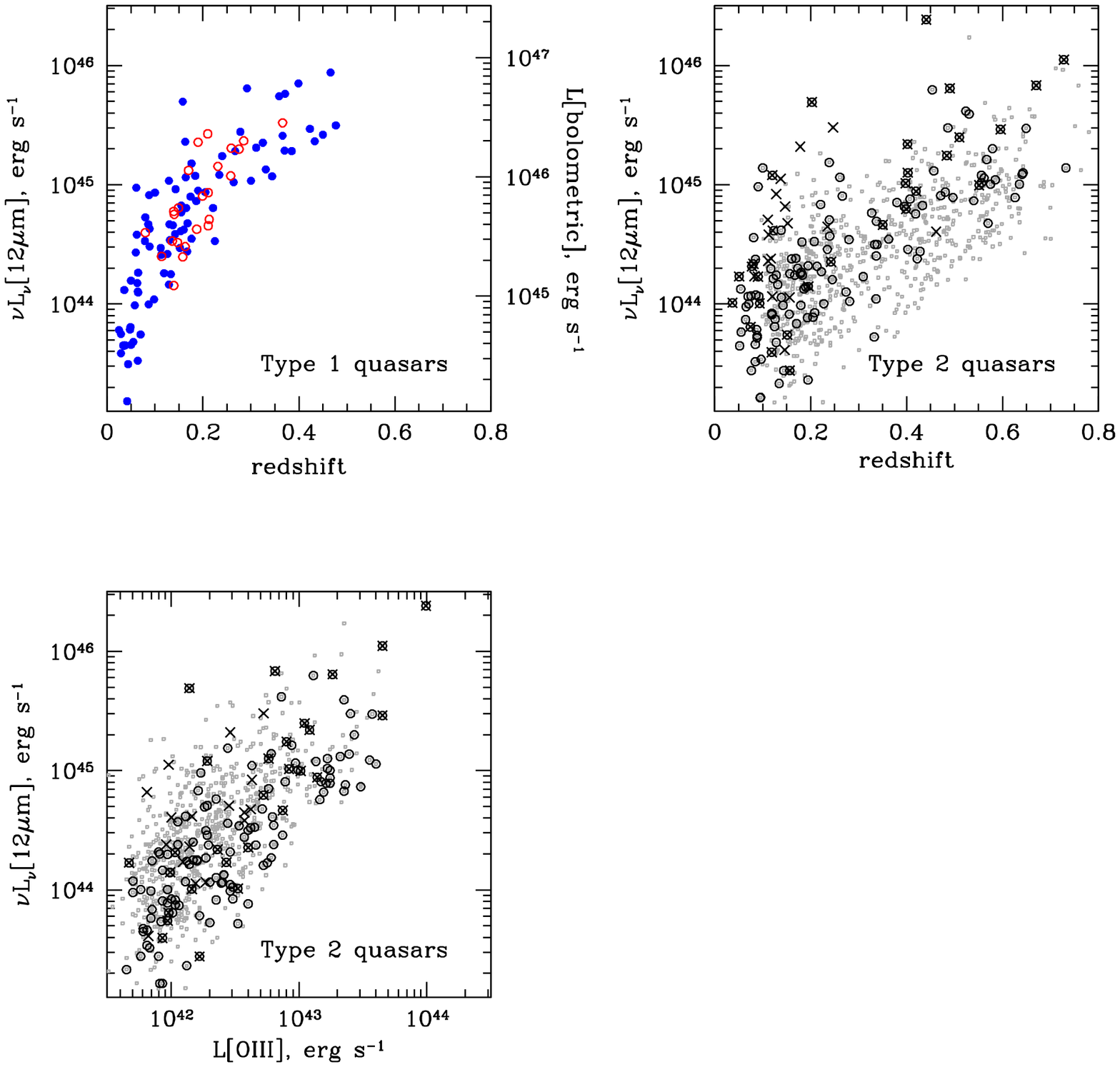}
\caption{[OIII] and 12\micron\ luminosities of type 2 quasars with 160\micron\ photometry from \spi\ or \her\ (circles; Section \ref{sec:photo_analysis}) and \spi\ spectroscopy (crosses; Section \ref{sec:spec_analysis}). The background grey points show the parent sample of type 2 quasars \citep{reye08}.}
\label{pic_dist_type2}
\end{figure}

For this sample, we collect archival \spi\ photometry and analyze new \her\ photometry as discussed in Sections \ref{sec:photo-spi} and \ref{sec:photo-her} for a total of 136 objects. Furthermore, while we previously published ten \spi\ spectra of type 2 quasars \citep{zaka08}, in Section \ref{sec:spec} we conduct an extensive archival search which allows us to significantly expand the sample and present 46 spectra here. The photometric and spectroscopic samples overlap by 28 objects. The distribution of [OIII] and mid-infrared luminosities for the parent sample and for the objects with follow-up \spi\ and \her\ observations is shown in Figure \ref{pic_dist_type2}. 

Our second sample is comprised of 115 type 1 quasars at $z\la 0.5$ studied with \spi\ spectroscopy by \citet{shi07}. Of these, 90 are ultraviolet-excess Palomar-Green (PG; \citealt{schm83, gree86}) quasars, and the remaining 25 are quasars selected from the Two Micron All Sky Survey (2MASS) with red $R-K$ colors \citep{cutr01, smit02}. Thus it is a heterogeneous quasar sample that includes objects with a range of extinction, from $A_V\simeq 0$ to $\la 5$ mag \citep{zaka05}, but overall it is dominated by type 1 (broad-line) sources. In the few cases of narrow-line (type 2) classification in the optical, broad emission lines and strong quasar continuum are seen in the near-infrared \citep{glik12}. Hereafter, we refer to these objects collectively as type 1 sources, sometimes making a distinction between `blue' and `red' as necessary, according to whether they are drawn from the PG sample or the 2MASS sample. 

Out of 115 type 1 quasars, all but one have complete 4-band photometry from {\it Wide-field Infrared Survey Explorer} (WISE; \citealt{wrig10}). We calculate rest-frame 12\micron\ mid-infrared luminosities by power-law-interpolating between WISE-3 and WISE-4 bands, and then estimate bolometric luminosities by applying bolometric correction of 8.6 from \citet{rich06}. \spi\ spectroscopy is available for all 115 objects \citep{shi07}. For red type 1 quasars, we collect archival \spi\ photometry (Section \ref{sec:photo-spi}) and for blue type 1 quasars we use recently published \her\ photometry (Section \ref{sec:photo-her}), so that 114 out of 115 objects have far-infrared photometric data. 

The redshifts and mid-infrared luminosity distributions of type 1 and type 2 samples are similar, as shown in Figure \ref{pic_dist}. For 39 of the 115 type 1 quasars, [OIII] luminosity measurements are available in the catalog of \citet{shen11}. For this subsample, we find $L_{\rm [OIII]}=10^{42.3\pm 0.6}$ erg s$^{-1}$ (average and standard deviation), similar to the range probed by the type 2 sample with follow-up infrared data ($10^{42.5\pm 0.5}$ erg s$^{-1}$).

\begin{figure*}
\centering
\includegraphics[scale=0.8, clip=true, trim=0cm 11cm 0cm 0cm]{picture_sfr15.eps}
\caption{Redshift and mid-infrared luminosity distributions of both quasar samples discussed in this paper. Left: PG (filled blue circles) and 2MASS (open red circles) type 1 quasars with \spi\ spectroscopy presented by \citet{shi07} and \spi\ and \her\ photometry. Right: type 2 quasars with 160\micron\ photometry from \spi\ or \her\ (circles) and \spi\ spectroscopy (crosses) and with parent sample \citep{reye08} in grey. We estimate the bolometric luminosities of type 1 quasars from their 12\micron\ monochromatic luminosities (right axis of the left panel) using \citet{rich06} bolometric correction of 8.6 (which ranges between 7.8 and 9.3 depending on the assumed spectral energy distribution). Type 2 quasars likely have higher bolometric corrections. }
\label{pic_dist}
\end{figure*}

Type 2 quasars are redder in the mid-infrared than type 1 quasars \citep{liu13b}. Specifically, the infrared power-law index between rest-frame 5 and 12\micron\ luminosities $\beta$ (defined as $\nu L_{\nu}\propto \lambda^{\beta}$) is $-0.05\pm 0.30$ (mean and standard deviation) for type 1 quasars from \citet{shi07}, whereas for the type 2 quasars from \citet{reye08} catalog it is $\beta=0.78\pm 0.45$. Furthermore, the ratio of 12\micron\ luminosities to $L_{\rm [OIII]}$ is 0.5 dex higher in type 1 quasars than in type 2s at the same emission line luminosity (Zakamska et al. in prep.). Both these factors suggest that 12\micron\ luminosity is not an isotropic measure of quasar luminosity and that type 2 quasars are obscured even at mid-infrared wavelengths. The bolometric corrections of type 2 quasars are therefore likely to be higher than those of type 1 quasars, perhaps by as much as a factor of $\sim 3$ (which would be necessary to reconcile the infrared-to-[OIII] ratios of type 1s and type 2s), but as they remain uncertain we do not show them in Figure \ref{pic_dist}.

\subsection{Far-infrared photometry with \spi}
\label{sec:photo-spi}

Composite spectral energy distribution models (e.g., \citealt{poll07, mull11b, chen15}) attempt to decompose emission from active galactic nuclei into a component powered by the black hole and a component powered by star formation in the host galaxy, and to use these measurements to determine the power of both processes. At the heart of these methods is the empirical notion that dust heated by the black hole accretion is warmer than that heated by starlight, and thus a quasar-dominated spectral energy distribution peaks at shorter wavelengths than that of a star-forming galaxy \citep{degr87}. Therefore, observing longward of thermal peaks maximizes sensitivity to star formation and minimizes contamination by the quasar. 

We cross-correlate the 887 type 2 quasars from \citet{reye08} against the \spi\ Heritage Archive and we find 62 distinct sources with nominal coverage at 160\micron\ (the longest available wavelength) by the Multiband Imaging Photometer for \spi\ (MIPS; \citealt{riek04}), and we download the corresponding 100 distinct Astronomical Observation Requests (AOR). We use filtered ({\sl mfilt, mfunc}) post basic calibrated data (PBCD) products to perform point-spread function (PSF) photometry. To this end, we generate PSF models using STinyTim (MIPS Instrument Handbook, 2011) and develop an analytic approximation to them using piece-wise Airy functions; a detailed description of the PSF is available in \citet{ania11}. 

With this in hand, we perform PSF photometry of the First Look Survey, which allows us to calibrate our PSF fitting procedure against the catalog of 160\micron\ sources by \citet{fray06}. We find that our measurements are systematically fainter than theirs by 24\%, which we attribute to color corrections which they applied and we did not. We take their fluxes to be `true' values and correct the systematic offset using a constant multiplicative factor (analogous to their use of color corrections), after which we find excellent agreement between their fluxes and ours within their stated absolute uncertainty of 25\%. In the absence of color information, we cannot tailor our color corrections to a specific target. Having thus calibrated our PSF photometry procedure, we apply it to the MIPS-160 data of type 2 quasars. 

Of the 62 sources, 11 have poor enough data quality (covered on the edges of big scans, gaps in coverage overlapping with the source location) that we do not consider them. The 51 sources with acceptable data quality are listed in Table \ref{tab:photo}; of these, 12 are detected, both as evaluated by the improvement in reduced $\chi^2$ over a continuum-only fit and by visual inspection. Following \citet{fray06}, we adopt 25\% as the photometric uncertainty. The median value of detected flux is 101 mJy. For the remaining sources we derive upper limits by fitting PSFs at multiple random locations within the field of the object and deriving the standard deviation of the fitted fluxes, which is taken as a 1$\sigma$ limit for point-source detection. In Table \ref{tab:photo}, we give 5$\sigma$ upper limits derived using this procedure. The median upper limit is 84 mJy. 

We then select all good observations of the 39 non-detected sources by choosing only those with the reported uncertainty in the vicinity of the object of $<0.4$ MJy/sr, which roughly corresponds to a 5$\sigma$ sensitivity for point source detection of 280 mJy. We then make cutouts from these data centered on the known positions of our sources and we stack them using error-weighted averaging. We find a strong ($\sim 10\sigma$) detection in the stacked image, with a PSF flux of 23 mJy, which we take to be an estimate of the mean flux of non-detected sources. We also conduct a null test, in which all images to be stacked are randomly offset by several pixels from the source position. There is no source detection in the null test stack. 

The sample of type 2 quasars with archival MIPS-160 data is heterogeneous, as described in Table \ref{tab:photo}. 25 objects constitute the full content of our targeted program (GO-3163, PI Strauss); they were selected based on [OIII]$\lambda$5007\AA\ luminosity ($L_{\rm [OIII]}\ge 10^{42.5}$ erg s$^{-1}$), tend to be at relatively high redshifts $(z\ga 0.30)$ and show low rates ($<20\%$) of MIPS-160\micron\ detection. Three were observed by other groups because they are powerful radio galaxies with strong enough line emission to make it into the [OIII]-selected sample of \citet{reye08}. Five objects at low redshifts were observed by other groups as candidate Ultraluminous Infrared Galaxies (ULIRGs) or type 2 quasars, and these are strongly detected with high fluxes. 18 objects are covered serendipitously by observations of other targets or calibration observations. Thus is it not surprising to have a few bright detections (in particular, nearby objects selected by other observers as ULIRG candidates) supplemented with many objects that are much fainter. 

For the 25 objects covered by our program GO-3163, we also have MIPS-70 measurements performed in 2006 (previously unpublished). For MIPS-70, we computed fluxes by aperture photometry using {\sl MOPEX} with aperture radius of $16''$ and applying aperture corrections derived from mosaicked images. The statistical errors were estimated from rms fluctuations of backgrounds. The color correction was applied assuming power-law flux density with the slope of $-1$. Twelve of the objects are detected at $>3\sigma$ level (whereas only four in the same program GO-3163 are detected in MIPS-160). In this paper we use these 12 detected sources to estimate infrared colors of type 2 quasars in Section \ref{sec:contrib}, leaving a detailed analysis of the spectral energy distributions for future. The MIPS-24 observations in this program have been superseded by WISE-4 data. 

As for the type 1 sample, the majority of blue quasars were observed by \her\ as described in the next section and in \citet{petr15}. Since \her\ data supersedes MIPS-160 data, we do not rematch the blue quasars to the \spi\ archive. All 25 red type 1 quasars are covered by archival MIPS-160 observations from two programs: 11 objects were observed by PI F. Low as a follow-up of 2MASS-selected quasars, and the remaining 14 objects were observed by PI G. Rieke as a follow-up of the most luminous quasars known at $z<0.3$. We analyze the photometry of these 25 sources in the same way as we do the type 2 sample and include them in Table \ref{tab:photo}. 17 objects are detected with a median flux of 139 mJy and for 8 objects we give upper limits with a median value of 114 mJy. 

\subsection{Far-infrared photometry from \her}
\label{sec:photo-her}

We proceed to \her\ photometry of type 2 quasars from \citet{reye08}. Our \her\ sample is assembled from two programs of pointed observations. In the first one (PI Zakamska), we obtained pointed observations of seven [OIII]-luminous sources ($L_{\rm [OIII]}\ge 10^{43.0}$ erg s$^{-1}$, median $L_{\rm [OIII]}=10^{43.2}$ erg s$^{-1}$) whose optical line emission was studied in detail by \citet{liu13a, liu13b}. In the second (PI Ho), we obtained pointed observations of 90 sources roughly matched in redshift, infrared luminosity and [OIII] luminosity to the PG sample (Figure \ref{pic_dist}) and sampling the full range of [OIII] luminosities ($L_{\rm [OIII]}=10^{41.7-43.4}$ erg s$^{-1}$) of the parent sample of \citet{reye08}. Similarly deep photometry was obtained in both programs using the Photodetector Array Camera and Spectrometer (PACS) in the mini-scan map mode at 70\micron\ and 160\micron\ and Spectral and Photometric Imaging Receiver (SPIRE) at 250\micron. All our targets are assumed to be point sources at \her\ resolution (the full width at half maximum of the point spread function is 12\arcsec).

For the smaller program (PI Zakamska), we use Level 2 PACS and SPIRE observations produced by standard pipeline reduction procedures (described in Chapter 7 of the PACS observing manual and in Chapter 5 of SPIRE data handbook). Source confusion is not an issue in PACS bands: at 0.7 mJy \citep{magn13}, confusion at 160\micron\ is well below our $1\sigma$ sensitivity of 2.5 mJy. We perform aperture photometry in the \her\ Interactive Processing Environment (HIPE) version 10.0 around the optical positions (known to better than 0.1\arcsec, with \her\ absolute pointing error of 0.81\arcsec, \citealt{sanc14}). We use the {\sl AnnularSkyAperturePhotometry} task within HIPE and apply aperture corrections using the {\sl PhotApertureCorrectionPointSource} task. We detect all seven sources at 70\micron\ and six of them at 160\micron\ at above 3$\sigma$, with median fluxes of 22 mJy and 16 mJy, respectively. This photometry is presented in Table \ref{tab:photo}. 

For the SPIRE images, we use an extraction and photometry task in HIPE that implemented the {\sl SUSSExtractor} algorithm described by \citet{sava07}. We do not detect any sources in the SPIRE bands, where our nominal $1\sigma$ point-source sensitivity is slightly below the confusion limit, 6 mJy at 250\micron\ \citep{nguy10}, and our images are indeed confusion-limited. Extrapolating our measured PACS-160 fluxes to the SPIRE-250 band using $F_{\nu}\propto \nu^{4.5}$ typical of the long-wavelength spectrum of star-forming galaxies \citep{kirk12}, we find that the median flux in SPIRE-250 is expected at the $\sim 3$ mJy level, below the confusion limit, thus the lack of detections is not surprising. 

\her\ data for type 2 quasars from the larger program (PI Ho) will be presented in their entirety by Petric et al. (in prep). Here we use exclusively 160\micron\ PACS fluxes from this program obtained using aperture photometry tools in HIPE in a manner similar to that described in \citet{petr15}. In this \her\ program, 90 objects were observed, 76 of them were detected and for the remaining 14 we set 4$\sigma$ upper limits. 

For blue type 1 quasars \her\ photometric data are published in \citet{petr15} and we use their 160\micron\ fluxes here. 85 objects were observed, 69 of them were detected and for the remaining 16 we use upper limits from \citet{petr15}. Out of the remaining five blue type 1s from the sample of \citet{shi07}, four have 160\micron\ photometry from \spi\ or {\it ISO} in the literature \citep{haas00, shan11}, and we include them in our analysis. 

\subsection{Spectroscopic observations}
\label{sec:spec}

Mid-infrared spectra of galaxies contain a wealth of information on star formation processes and on the nuclear activity, including the emission features of polycyclic aromatic hydrocarbons (PAHs; \citealt{alla89, rous01, dale02}) and the low-ionization and high-ionization ionic emission lines ([NeVI]$\lambda$7.65\micron, [SIV]$\lambda$10.51\micron, [NeII]$\lambda$12.81\micron, [NeV]$\lambda$14.32\micron, [NeIII]$\lambda$15.56\micron, \citealt{farr07, inam13}). Our analysis is based on the \spi\ Space Telescope Infrared Spectrograph (IRS; \citealt{houc04}) low-resolution spectra of quasars of different types. For type 1 blue PG quasars and red 2MASS quasars, we use published spectra and analysis by \citet{shi07}. As for type 2 quasars, we cross-correlate the type 2 quasar sample \citep{reye08} against IRS data using \spi\ Heritage Archive. 

\begin{figure}
\centering
\includegraphics[scale=0.45, clip=true, trim=0cm 7cm 0cm 0cm]{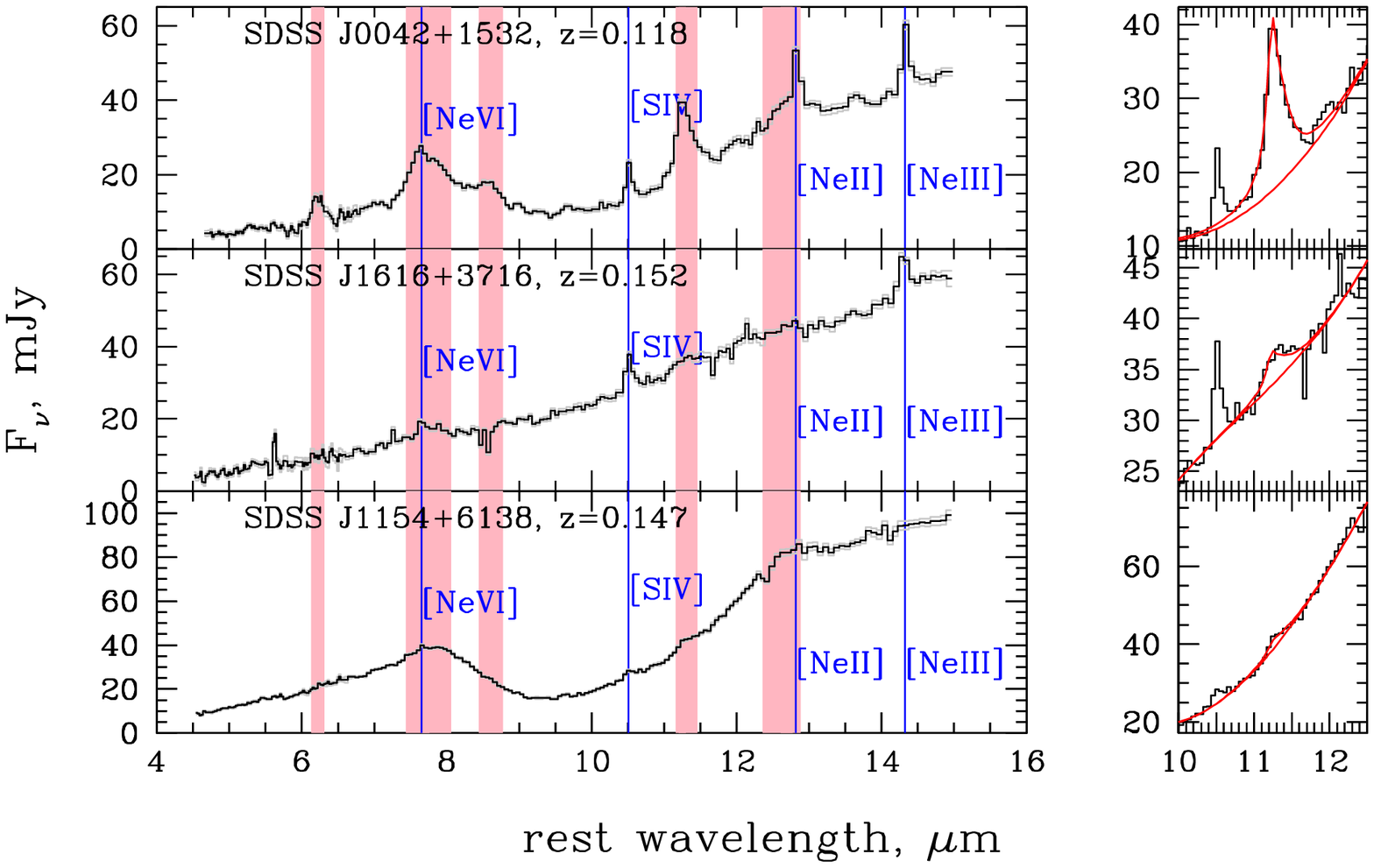}
\caption{Example spectra of three type 2 quasars from our sample: from top to bottom, a spectrum with relatively strong PAHs, a power-law-dominated spectrum and a silicate-absorption-dominated spectrum. On the right, we show PAH[11.3\micron] fits with fixed shape of the PAH feature taken from \citet{smit07} and third order polynomial continuum. Pink shading shows the expected locations of PAH complexes, vertical blue lines show positions of some of the brightest emission lines. The broad absorption feature extending from $\sim 8$\micron\ to $\sim 13$\micron\ in SDSS J1154+6138 is due to silicates.}
\label{pic_example}
\end{figure}

We find 46 type 2 quasars from \citet{reye08} with IRS spectra of varying quality within 3\arcsec\ of the optical position. In Table \ref{tab:spec} we list type 2 quasars with mid-infrared spectroscopic measurements as well as comments on how these objects were selected for follow-up spectroscopy. The majority were targeted by various groups as type 2 quasar candidates. Ten of them were from our own program \citep{zaka08} and were selected based on [OIII] luminosity and infrared flux ($L_{\rm [OIII]}>10^{42.6}$ erg s$^{-1}$, $F_{\nu}$[8\micron]$>1.5$ mJy, $F_{\nu}$[24\micron]$>6$ mJy). Other programs selected targets based on X-ray properties and optical or infrared luminosity diagnostics. Thus the sample is a fairly representative subsample of the \citet{reye08} sample of type 2 quasars (Figure \ref{pic_dist}). Depending on the redshifts of the targets and on which IRS gratings were used for the observations, the wavelength coverage ranges from $\sim$5-13\micron\ in the rest-frame (12 objects) to $\sim$5-25\micron\ (the rest of the sample). Example spectra are shown in Figure \ref{pic_example}. There are 28 objects in common between the type 2 sample with IRS spectra and the type 2 sample with 160\micron\ photometric data; these sources are discussed in Section \ref{sec:effect}. 

As the majority of the sources are point-like at \spi\ resolution, we use the enhanced data products described in Chapter 9 of the IRS Instrument Handbook. In a handful of cases where several spectra of the same target are returned by the search engine (perhaps because the IRS coordinates are slightly offset from one another) we combine the spectra into one using error-weighting. We inspect all spectra to make sure that the short- and the long-wavelength (SL and LL) spectra stitch together well in the region of the overlap. Because the LL grating has a larger aperture than the SL one, for extended sources some of the flux may be missed by the SL grating \citep{bran06}; furthermore, even for point sources slight relative mis-alignment of the gratings would result in a greater loss of flux from the SL slit. In 10 cases we apply a multiplicative factor $>$1 to the SL spectrum to bring it into the agreement with the LL spectrum; only in 5 of these cases is the adjustment greater than 10\%. 

With these spectra in hand, we double-check their absolute flux calibrations. Because we primarily use PAH[11.3\micron] fluxes in the analysis which follows, absolute flux calibration around this wavelength is particularly important. We convolve the spectra with the Wide-Field Infrared Survey Explorer (WISE) filter curves from \citet{jarr11}, obtain synthetic fluxes in the WISE-3 band (effective wavelength 11.6\micron) and compare those with observed WISE-3 fluxes. They show excellent agreement, with the average ratio between synthetic fluxes and observed fluxes of 0.03 dex and the standard deviation among the 46 objects of 0.04 dex. We therefore take 0.05 dex (12\%) to be the absolute flux calibration uncertainty for these sources. 

To calculate PAH fluxes, we cut out $\la$3\micron-wide chunks of the spectrum and model them using a polynomial continuum and Drude profiles with profile shapes and widths taken from \citet{smit07}. Drude (or damped harmonic oscillator) profiles arise in the Drude theory of conductivity \citep{bohr83} and are found to be very well matched to ultra-violet and infrared opacity curves of small dust particles \citep{fitz86}. For PAH complexes, such as those at 11.3\micron\ and at 7.7\micron, the relative amplitudes of the components within the complex are fixed to their ratios in the template spectrum of normal star-forming galaxies \citep{smit07}. For example, within the 11.3\micron\ complex the amplitude ratio of the 11.23\micron\ and 11.33\micron\ components is fixed to 1.25:1. Depending on the model for the local continuum, from a constant to a cubic polynomial, the number of fit parameters varies from two to five, with the amplitude being the only parameter that describes the intensity of the PAH feature (since the functional shape of the feature remains fixed). Example fits are shown in Figure \ref{pic_example}. Overall the 11.3\micron\ and the 6.2\micron\ features are reproduced well, but the quality of fits of the 7.7\micron\ feature is poor. Contributing factors are a strong [NeVI]7.652\micron\ emission line blended with the PAH complex and a poorly anchored continuum, whose shape is complicated by silicate absorption. We therefore do not use the results from the PAH[7.7\micron] fits. 

The mid-infrared continua of obscured quasars show a wide range of behavior of the silicate feature centered at 9.7\micron, from deep absorption to occasional emission \citep{stur06, zaka08}. We measure the apparent strength of silicate absorption defined as $S[9.7\micron]=-\ln(f_{\rm obs}[9.7\micron]/f_{\rm cont}[9.7\micron])$, where $f_{\rm obs}$ is the observed flux density at 9.7\micron\ and $f_{\rm cont}$ is the estimate of silicate-free continuum obtained by power-law interpolation between $5.3-5.6\micron$ and $13.85-14.15\micron$. Negative values of $S[9.7\micron]$ indicate silicate emission, while positive values indicate absorption, with $S[9.7\micron]\ga 1$ for the 10\% most absorbed sources. This method is similar to that used by \citet{spoo07}, except we do not use a continuum point at 7.7\micron\ even in the cases of weak PAH emission. The apparent strength of Si absorption is closely related to, but not identical to the optical depth of Si dust absorption; depending on the poorly known continuum opacity of the dust at these wavelengths, the actual optical depth is $\simeq (1.2-1.5)\times$ the apparent strength of the Si feature \citep{zaka10}. 

Forbidden lines [NeII]$\lambda$12.81\micron, [NeIII]$\lambda$15.56\micron, [NeV]$\lambda$14.322\micron, [NeVI]$\lambda$7.652\micron\ and [SIV]10.51\micron\ are measured by fitting Gaussian profiles plus an underlying linear continuum. As these lines are not spectrally resolved, their widths $\sigma$ in the observer's frame are fixed to the order-dependent instrumental resolution tabulated by \citet{smit07}. We cut out a $3\sigma-$wide piece of spectrum centered on the emission line in question and perform a three-parameter fit, with two parameters describing the continuum and one parameter for the line amplitude. We allow for an 0.03\micron\ variation in the line centroid to account for the wavelength calibration uncertainty \citep{smit07}. Because our fits for PAH emission features and forbidden emission lines are linear in all parameters, we use the standard error as the estimate of the standard deviation of the parameter estimate. 

In addition to the 46 type 2 quasars with spectra, we use 115 IRS spectra for all type 1 quasars from \citet{shi07}. As was described in the beginning of Section \ref{sec:data}, 90 of these are optically selected blue PG quasars and 25 are near-infrared-selected quasars of varying optical types. The IRS sample was assembled from several dedicated programs and archival search as described by \citet{shi07}, and their spectra were analyzed in detail using methods similar to ours. In each of the three subsamples (blue type 1 quasars, red type 1 quasars, type 2 quasars) the detection rate of the 11.3\micron\ PAH feature is $\sim$ 50\%, and we use upper limits on PAH fluxes in the remaining objects. 

\subsection{Radio data}
 
We cross-match all objects with spectroscopic or photometric infrared data with the Faint Images of Radio Sky at Twenty cm survey (FIRST; \citealt{beck95, whit97}) within 3\arcsec\ of the optical position. FIRST used the Very Large Array to produce a catalog of the radio sky at 1.4 GHz with a resolution of 5\arcsec, subarcsec positional accuracy, rms sensitivity of 0.15 mJy and catalog threshold of $\sim 1.0$ mJy for point sources. When a source is covered by the FIRST data but there is no catalog detection, we estimate the flux density upper limit as 5$\times$rms flux density at source position$+0.25$ mJy, with the last term included to correct for the CLEAN bias \citep{whit97}. 

In cases of no FIRST coverage (7\% of type 2 quasars and 25\% of type 1 quasars), we use the NRAO VLA Sky Survey (NVSS; \citealt{cond98}), which is a 1.4 GHz survey covering the entire sky north of $-40\deg$ with a resolution of 45\arcsec, positional accuracy of better than 7\arcsec, rms sensitivity of $\sim 0.4$ mJy and catalog threshold of $\sim 2.3$ mJy. We match within 15\arcsec\ of the optical position and in case of non-detections, calculate the upper limit as 5$\times$rms flux density at source position$+0.4$ mJy \citep{whit97}. 

Our matching procedure implies that for extended radio sources -- a minority of our sample -- we are sensitive only to the core fluxes, not to the extended lobes. Inclusion of lobe emission would increase the observed radio luminosities quoted in this paper, but only for a minority of sources. Most objects are point-like at the resolution of FIRST and NVSS \citep{zaka04}, with only 10\%-20\% sources (both in the type 2 and the type 1 samples) showing integrated fluxes significantly higher than peak fluxes. The radio detection rates are 73\% for the type 2 sample with far-infrared photometry, 85\% for the type 2 sample with IRS spectroscopy, and 59\% for the type 1 sample.

All radio luminosities quoted in this paper are K-corrected to rest-frame 1.4 GHz using equation 
\begin{equation}
L_{\rm radio,obs}\equiv\nu L_{\nu}[1.4{\rm GHz}] =4\pi D_L^2 \nu F_{\nu}(1+z)^{-1-\alpha},
\end{equation}
where $\nu=1.4$ GHz, $D_L$ is the luminosity distance, $F_{\nu}$ is the observed flux density at 1.4GHz, and $\alpha$ is the power-law spectral index defined as $F_{\nu}\propto \nu^{\alpha}$. Radio-quiet quasars at $z\sim 0.5$ are too faint to be detectable by any large radio surveys other than FIRST and NVSS, which have data only at 1.4 GHz, so we cannot measure $\alpha$ from archival data. Values between $-0.5$ and $-1.0$ for the radio-quiet population were suggested in the literature \citep{barv89, ivez04, zaka04}; unless specified otherwise, we assume $\alpha=-0.7$. For a source at $z=0.5$ with a fixed observed flux density $F_{\nu}$, varying $\alpha$ from -0.7 in its typical range between -0.5 and -1 results in a 10\% uncertainty in $L_{\rm radio,obs}$. 

\section{Star formation rates of quasar hosts from photometry}
\label{sec:photo_analysis}

Dust that produces infrared emission of quasars and star forming galaxies is heated by the radiation from the accretion disk or from young stars. Because of the very high optical depths involved, all incoming radiation at optical and ultraviolet wavelengths is absorbed in a thin layer close to the source of the emission and then thermally reprocessed thereafter. Therefore, it is unlikely that any differences between radiation fields in active and star forming galaxies may be responsible for the noticeable differences in the infrared spectral energy distributions. 

Instead, the biggest difference between quasar-heated and star-formation-heated dust is that the latter is distributed over the entire galaxy on $D_{\rm gal}\ga 1$ kpc scales, whereas dust heated by an AGN is concentrated on scales $D_{\rm qso}\la 10$ pc even in luminous objects \citep{kish11}. A star-forming galaxy and an active nucleus of similar bolometric luminosities ($L\propto D^2 T^4$) would have different characteristic dust temperatures, $T_{\rm gal}/T_{\rm qso}\sim \sqrt{D_{\rm qso}/D_{\rm gal}}\sim 0.1$. This crude scaling is borne out by far-infrared observations of star-forming galaxies, whose characteristic temperature is $T_{\rm gal} \simeq 25$ K, and of AGN, where the bulk of the thermal emission is produced with $T_{\rm qso}\gg 100$ K \citep{rich06, kirk12}. Beyond this basic temperature distinction, a variety of shapes of the spectral energy distributions can be produced due to the differences in the geometric distribution of dust (compact vs diffuse, spherical vs non-spherical, clumpy vs non-clumpy, etc.), its amount, and its orientation relative to the observer \citep{pier92, nenk02, leve07}. 

Because of the steep decline of the modified black body function at wavelengths greater than those that correspond to the thermal peak, in composite sources with similar contributions from the active nucleus and the star forming host galaxy the mid-infrared emission tends to be dominated by the active nucleus and the far-infrared emission ($\lambda\ga 100\micron$) is dominated by star formation \citep{hatz10}. But in quasars even the longest wavelength emission probed by \her\ observations can be dominated by emission from hot (presumably quasar-heated) dust \citep{hony11, sun14}. 

Our approach is therefore to use far-infrared observations to obtain strict upper limits on the quasar hosts' star formation rates. To minimize the contribution from the quasar -- insofar as it is possible -- we use the longest wavelength observations available to us. In practice, we use 160\micron\ data either from \spi\ or from \her. We then assume that all of the observed far-infrared emission is due to star formation, and calculate the corresponding star formation rates and the expected radio luminosities \citep{helo85, bell03, mori10}. This predicted radio luminosity is an upper limit on the amount of radio emission that can be generated by star formation. We then compare these predictions with the observed radio emission. By using a variety of templates to estimate star formation rates, we ensure that our results are robust to varying the assumed spectral energy distribution of a star-forming galaxy. 

Finally, our measurements are predicated on the assumption that the far-infrared fluxes in star-forming galaxies are measuring the instantaneous rates of star formation. This is not always true \citep{hayw14}, in that previously formed stars can continue to illuminate left-over dust even after star formation rates have declined. But because this effect results in an over-estimate of star formation rate when using far-infrared fluxes, it is consistent with our upper-limit approach. 

\begin{figure*}
\centering
\includegraphics[scale=0.7, clip=true, trim=0cm 10cm 10cm 0cm]{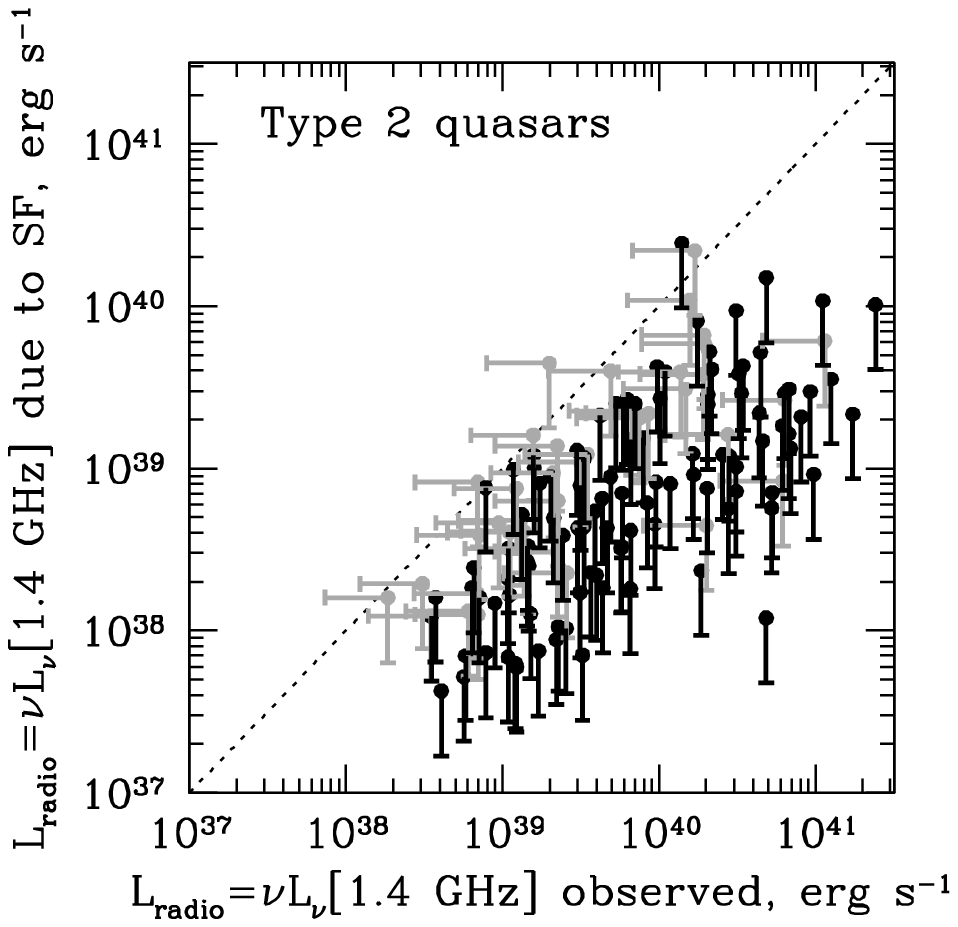}%
\includegraphics[scale=0.7, clip=true, trim=0cm 10cm 10cm 0cm]{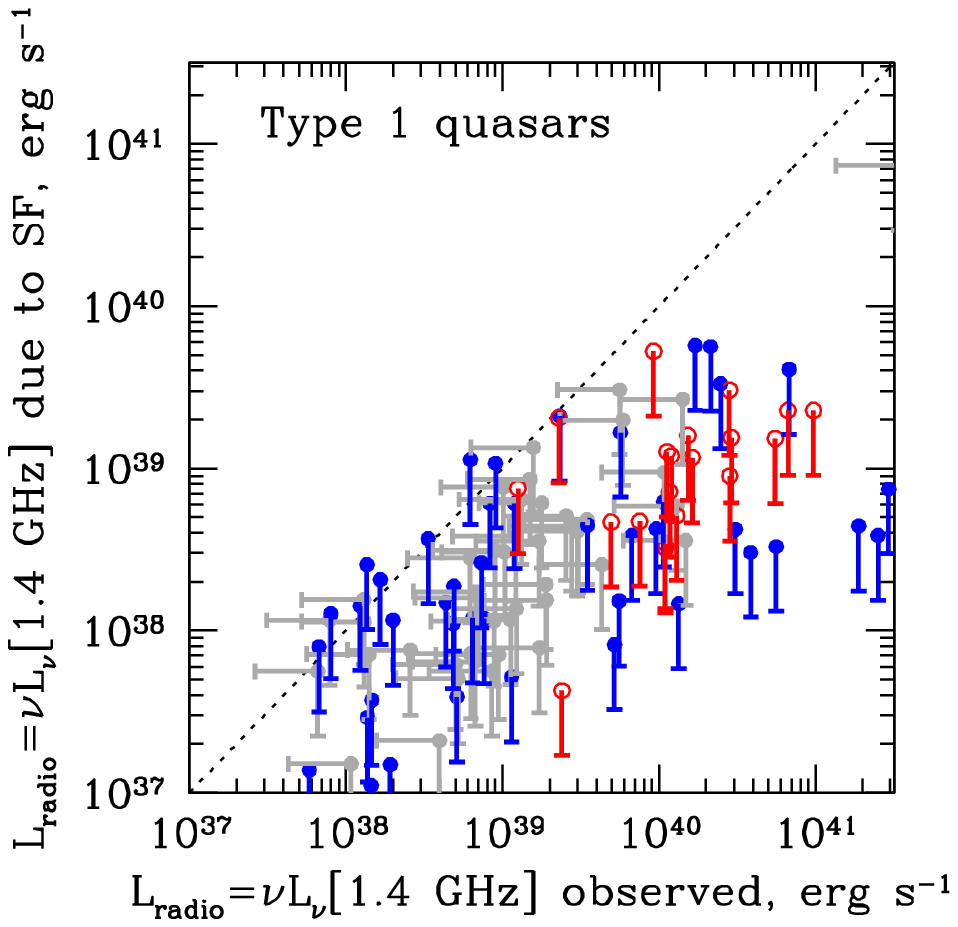}\\
\caption{Left: Results from the far-infrared photometric data from \spi\ and \her\ 160\micron\ observations of type 2 quasars from \citet{reye08}. We plot radio luminosity expected due to star formation in the quasar hosts vs the observed radio luminosity. Four points corresponding to radio-loud sources with $L_{\rm radio,obs}>3\times 10^{41}$ erg s$^{-1}$ and $L_{\rm radio,SF}<10^{41}$ erg s$^{-1}$ are off the scale of the plot to the right. Vertical bars show that all points represent upper limits on host star formation rates (and thus upper limits on the associated expected radio luminosity). Grey points with horizontal error bars denote points that are not detected by FIRST / NVSS, whereas black symbols correspond to radio detections. In the cases where radio emission is detected (74\% of the objects), star formation is insufficient to account for the observed radio emission, with the median $\log(L_{\rm radio,obs}/L_{\rm radio,SF})=1.0$. Right: Same calculation for type 1 quasars. Blue points are for PG quasars (predominantly \her\ data from \citealt{petr15}) and red for 2MASS quasars (MIPS-160 data) detected in the radio, and grey points are for radio non-detections. For these objects, the median $\log(L_{\rm radio,obs}/L_{\rm radio,SF})=1.1$. Uncertainties in radio fluxes as less than 15\%.} 
\label{pic_photo}
\end{figure*}

The main result of this section is presented in Figure \ref{pic_photo} which demonstrates that the radio emission due to star formation in the quasar hosts is inadequate -- by almost an order of magnitude -- to explain the observed radio emission. Below we describe in detail the steps involved in this comparison and in the cross-checking of this result we performed using a variety of methods. 

\subsection{The infrared-radio correlation of star-forming galaxies}
\label{sec:ir_radio}

The key to making an accurate comparison between the observed radio luminosity and that predicted from star formation in the host galaxy is a careful calibration between the far-infrared luminosities, star formation rates and radio luminosities due to star formation in star-forming galaxies without an active black hole. The strong correlation between these values is due to massive young stars which dominate the ultraviolet continuum most easily absorbed by interstellar dust, resulting in a `calorimetric measure' of star formation rates \citep{kenn98}. The same young stars explode as supernovae, resulting in acceleration of cosmic rays which produce the observed radio emission \citep{helo85}. 

The tightest correlation is between the total infrared luminosity of star formation (by convention, often integrated between 8 and 1000\micron, $L_{\rm 8-1000\micron, SF}\equiv L_{\rm IR,SF}$) and radio luminosity, which we take from Figure 3 of \citet{bell03}: 
\begin{equation}
\log L_{\rm radio,SF}{\rm [erg\, s^{-1}]}=26.4687+1.1054 \log (L_{\rm IR, SF}/L_{\odot}).\label{eq_ir_radio}
\end{equation} 
In this equation and hereafter, the radio luminosity is the monochromatic luminosity at rest-frame 1.4 GHz, $L_{\rm radio}\equiv \nu L_{\nu}$[1.4GHz].

To reassure ourselves that this relationship applies to galaxies with a wide range of star formation rates -- including the high star formation rates we confront in infrared-luminous sources discussed in this paper -- we double-check this conversion against data for two samples of star-forming galaxies analyzed completely independently from \citet{bell03} by several different groups. At low luminosities, we use nearby galaxies from \citet{mull11b}. At high luminosities, we use the Great Observatories All-Sky Luminous Infrared Galaxy Survey (GOALS; \citealt{armu09}). In both cases, we take advantage of the 8-1000\micron\ infrared luminosities tabulated by \citet{mull11b} and \citet{armu09} and obtain radio luminosities from the NVSS. 

Because the GOALS sample may contain active galactic nuclei, we restrict our comparison to those objects that have mid-infrared classifications consistent with pure star formation, by requiring the rest-frame equivalent width (EW) of the polycyclic aromatic hydrocarbon emission at 6.2\micron\ to be above 0.3\micron\ \citep{stie13}. Although not a perfect diagnostic, this measure is reasonably well correlated with ionization-line diagnostics of AGN activity \citep{petr11}. Furthermore, many GOALS galaxies are found in mergers, and the total infrared luminosities of these sources include all components found within the $\sim 5\arcmin$ beam of the Infrared Astronomical Satellite \citep{neug84} which is significantly larger than the NVSS beam ($\sim 45\arcsec$). Thus, in widely separated mergers the infrared emission could include multiple interacting components, while the radio emission would be coming from only one of them. To make sure we compare fluxes from similar apertures, we further restrict our comparison to objects that are either in single non-interacting hosts or in late-stage mergers (`N' and `d' classifications of \citealt{stie13}), excluding pairs and triples at all interaction stages. For these objects, using NVSS fluxes ensures that the total radio emission is taken into account. 

Overall we find good agreement between the infrared-radio correlation reported by \citet{bell03} and that displayed by these two samples of star-forming galaxies which sample three orders of magnitude in infrared luminosity. Given a measurement of the total infrared luminosity, the standard deviation of the radio luminosity of two samples around the best-fit correlation is 0.16 dex (for \citealt{mull11b} galaxies) and 0.24 dex (for GOALS galaxies), which we take to be the practical measure of the uncertainty in the $L_{8-1000\micron}$-radio correlation. 

In quasars, the total infrared luminosity may be dominated by the activity in the nucleus, and without many additional assumptions we cannot obtain a measurement of total infrared luminosity due to star formation alone. Thus our challenge is to make the best guess of the upper limit on the star formation rate and on the associated radio emission from just one photometric datapoint at 160\micron. To this end we use the calibrations presented by \citet{syme08} for star forming galaxies derived from deep MIPS data:
\begin{eqnarray}
\log (L_{\rm IR, SF}/L_{\odot}) = 1.16 + 0.92 \log (\nu L_{\nu}[70\micron]/L_{\odot});\label{eq_sym1}\\
\log (L_{\rm IR, SF}/L_{\odot}) = 1.49 + 0.90 \log (\nu L_{\nu}[160\micron]/L_{\odot}).\label{eq_sym2}
\end{eqnarray}
These relationships are well calibrated even in the highly star-forming regime.

\begin{figure}
\centering
\includegraphics[scale=0.6, clip=true, trim=0cm 10cm 10cm 0cm]{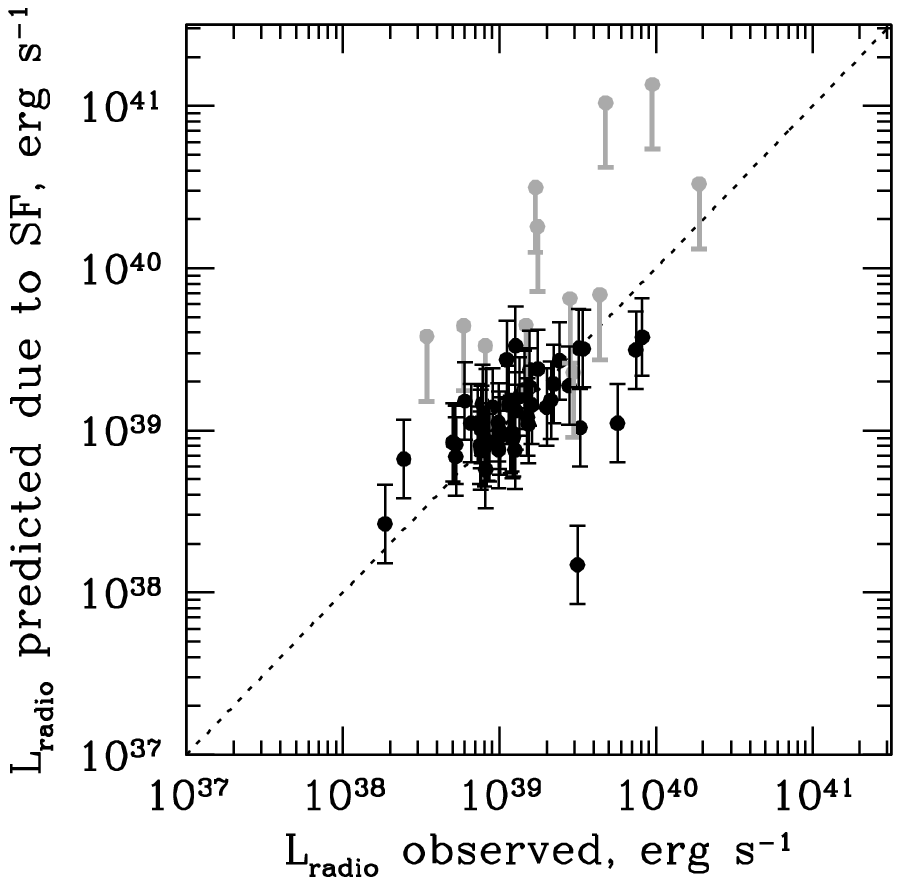}
\caption{Comparison between predicted radio emission due to star formation obtained from 160\micron\ fluxes from scaling relationships (\ref{eq_sym2}) and (\ref{eq_ir_radio}) and the observed radio emission for GOALS galaxies \citep{armu09, u12}, most of which are dominated by star formation. Black points show 160\micron\ detections, with 0.24 dex error bars which reflect the conversion of total infrared luminosity to predicted radio due to star formation for GOALS galaxies and grey points show 160\micron\ upper limits. We find excellent agreement between the observed and predicted radio luminosities in galaxies with 160\micron\ detections, with a standard deviation around the 1:1 relationship (dotted line) of 0.18 dex and a mean difference of 0.03 dex. The three most significant outliers below the dotted line all contain active nuclei, which presumably contribute excess radio emission over that associated with star formation alone.}
\label{pic_symcheck}
\end{figure}

Thus our calculation of the expected radio emission due to star formation involves the following steps. We use equations (\ref{eq_sym1})-(\ref{eq_sym2}) to convert a far-infrared photometric detection to the total luminosity of star formation, and then we use equation (\ref{eq_ir_radio}) to derive the expected radio luminosity. To verify that the scaling relationships give the correct answer for star forming galaxies, we apply this method to the GOALS sample in Figure \ref{pic_symcheck}. We use 160\micron\ photometry from \citet{u12}. Since these authors concentrate on the nearby ($z<0.083$) subsample, we can use equation (\ref{eq_sym2}) directly without any need for k-corrections. 

We find an excellent agreement between the observed radio luminosity and that predicted from the 160\micron\ flux via the scaling relationships. Only three points show a significant ($>0.4$ dex) excess of radio emission over that predicted from the 160\micron\ fluxes, and all three turn out to contain luminous active nuclei (MCG-03-34-064 is a Seyfert 1 galaxy, and NGC 5256 and NGC 7674 are Seyfert 2s, \citealt{petr11}) which likely contribute radio emission in excess of that due to star formation in the host. Excluding these three sources, we find that the median / average ratio of the observed-to-predicted flux is 0.01 / 0.03 dex, and the standard deviation around the 1:1 relationship is 0.18 dex. Therefore, we assume that the typical uncertainty in our method of predicting radio emission due to star formation from 160\micron\ band fluxes is about 0.2 dex, which is the combination of the standard deviation around the correlation and the typical photometric error of 160\micron\ observations (20-25\%, or $<0.1$ dex). Encouraged by such excellent agreement between predicted and observed radio fluxes in luminous star-forming galaxies, we apply the same method to quasars in the next section. 

\subsection{Radio emission in quasars is not due to star formation}
\label{sec:rem_photo}

We use the observed 160\micron\ fluxes (or upper limits) of the quasars in our samples to derive the upper limit on their total infrared luminosity due to star formation using equations (\ref{eq_sym1})-(\ref{eq_sym2}). Because the relations are given at rest-frame 70\micron\ and 160\micron, and the spectral slopes of our targets are unknown, instead of performing k-corrections on the data we linearly interpolate the slopes and the normalizations of equations (\ref{eq_sym1})-(\ref{eq_sym2}) between rest-frame 70\micron\ and 160\micron\ depending on the redshift of each target, thereby establishing a relation between monochromatic luminosity and the total luminosity of star formation at rest-frame wavelength of 160\micron/$(1+z)$. We then use equation (\ref{eq_ir_radio}) to derive an upper limit on the radio emission due to star formation and compare with the observed amount. 

The results for both type 1 and type 2 quasar samples are shown in Figure \ref{pic_photo}. Unlike GOALS galaxies in Figure \ref{pic_symcheck}, almost all quasars in our sample show significantly higher radio luminosities than those expected from star formation (and furthermore the predicted radio emission is a strict upper limit on the star formation contribution, as reflected in the one-sided error bars, as not all of the 160\micron\ continuum is due to star formation). Among the quasars which have strong detections in the radio, the median ratio between observed radio luminosity and that expected from star formation is an order of magnitude: $\log (L_{\rm radio,obs}/L_{\rm radio,SF})=1.1$ for type 1 quasars and $1.0$ for type 2s. The standard deviation of this ratio is 0.7 dex for type 2 quasars and 1.1 dex for type 1s, though in the latter case the ratio is not log-normally distributed and the standard deviation is artificially inflated by eight radio-loud sources with $L_{\rm radio}\ge 10^{41}$ erg s$^{-1}$. Removing these, we find a standard deviation of 0.7 dex for type 1s as well. In any case, the standard deviation is much greater than the $\sim$ 0.2 dex standard deviation in the calibrations of star formation rates in star-forming galaxies and the $\sim 0.1$ dex 160\micron\ flux uncertainty for detections.

Our main conclusion is that the star formation in quasar hosts falls short of explaining the observed radio emission in quasars by about an order of magnitude. This is consistent with the study by \citet{harr14} who found, using the spectral energy decomposition methods, that the observed radio luminosities were in all cases well above the calculated star formation component in their sample. Quasars in their sample are somewhat less luminous than ours, so the effect is likely more pronounced in our case: at the same star formation rate \citep{stan15}, an increase in the quasar contribution would make the radio/160\micron\ ratio more discrepant from that measured in star forming galaxies.  

It is more difficult to draw conclusions from the radio non-detections in the FIRST and / or NVSS surveys (shown as grey points in Figure \ref{pic_photo}), because in this case both the predicted radio luminosity due to star formation and the observed luminosity are only available as upper limits. Nonetheless, these objects do not alter our main conclusion. We have stacked the FIRST images of the non-detected sources in Figure \ref{pic_photo}, left, and obtained a strong point source detection with a mean peak flux of 0.4 mJy/beam. This estimate is in excellent agreement with our previous finding in Stripe 82 \citep{zaka14}, where we were able to detect all FIRST-undetected sources in a more sensitive survey \citep{hodg11} with fluxes about a factor of 2 below the limit of the FIRST survey ($\sim 1$ mJy). If in Figure \ref{pic_photo} all sources without radio detections have typical fluxes of 0.4 mJy, then we can again calculate the excess of observed radio emission over the upper limit on radio emission from star formation, which we find to be $\log (L_{\rm radio,obs}/L_{\rm radio,SF})=0.59$ for type 1 quasars and $=0.74$ for type 2s.

To make sure that our results are robust to changes in the spectral energy distribution of star formation, we use several star formation templates available in the literature to recalculate the total infrared luminosity of star formation. We use seven templates, five from nearby star forming galaxies by \citet{mull11b} and two from $z=1-2$ star forming galaxies by \citet{kirk12}. We scale the templates (properly adjusted for redshift) to reproduce the observed 160\micron\ fluxes of our sources, with one fitting parameter -- the overall luminosity of the template. For each template, we obtain the total infrared luminosity $L_{\rm IR, SF}$ due to star formation by integrating the scaled template between 8 and 1000\micron, and of the seven results we pick the highest one, in keeping with the strict upper limits approach, which we convert to the expected radio luminosity \citep{bell03}. The results are qualitatively similar to those obtained via scaling relations from \citet{syme08} and shown in Figure \ref{pic_photo}, though the star formation rates obtained using the template method are systematically higher by about 0.2 dex. The reason for this is that we pick the most conservative template -- the one that gives us the highest star formation rate at a given 160\micron\ flux. Even with this method, the observed radio luminosities of quasars are in excess of those predicted from star formation, with $\log (L_{\rm radio,obs}/L_{\rm radio,SF})=0.6$.

\subsection{Contribution of the active nucleus to the far-infrared flux}
\label{sec:contrib}

\begin{figure}
\centering
\includegraphics[scale=0.45, clip=true, trim=0cm 10cm 0cm 0cm]{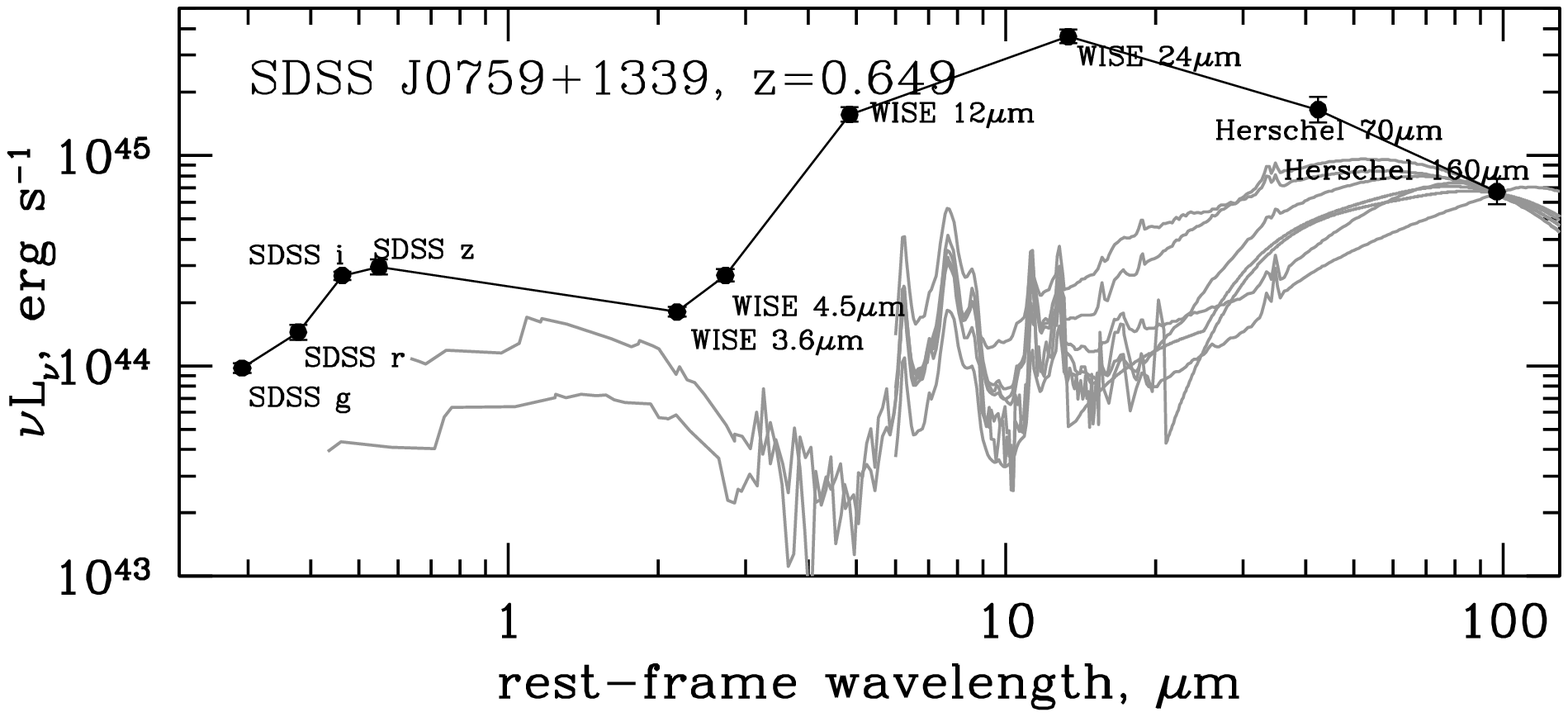}
\caption{The spectral energy distribution of one of the obscured quasars from \citet{reye08} and \citet{liu13a}. Seven star-formation templates from \citet{kirk12} and \citet{mull11b} are scaled to match the longest wavelength point. While it is clear that the overall spectral energy distribution of the object is inconsistent with any of the star formation templates (with excess emission at mid-infrared wavelengths likely due to quasar-heated dust), we can use the longest wavelength detection to place strict upper limits on the star formation rate in the quasar host galaxy. }
\label{pic_temp}
\end{figure}

Figure \ref{pic_temp} shows the comparison between the spectral energy distribution of one of our obscured quasars and the star-forming galaxy templates. The spectral energy distribution is assembled from photometric data from SDSS, WISE and \her, and the seven star formation templates \citep{mull11b, kirk12} are scaled to match the 160\micron\ observation. The spectral energy distribution of this object peaks at significantly shorter wavelengths (between 10 and 20\micron) than that of any of the star formation templates (between 50 and 150\micron); this is typical of our targets. 

\begin{figure*}
\centering
\includegraphics[scale=0.8, clip=true, trim=0cm 10cm 0cm 0cm]{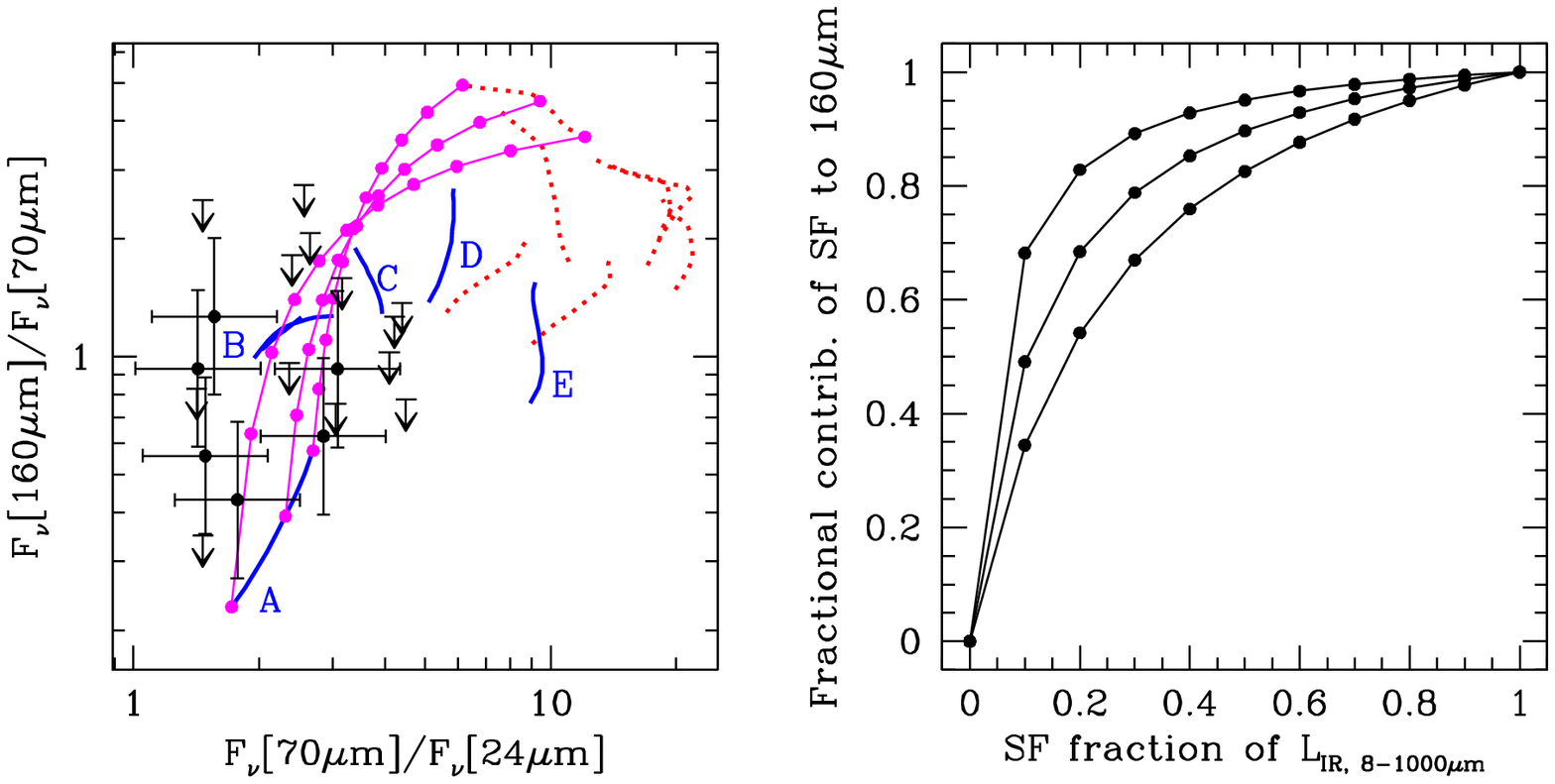}
\caption{Left: Infrared colors of type 2 quasars: black points and upper limits for \her\ and \spi\ observations of type 2 quasars detected at 70\micron\ in our targeted deep \spi\ observations. Dotted red lines show infrared colors of seven star formation templates \citep{mull11b, kirk12} placed at $z=0.24-0.73$. Solid blue lines show the same for six obscured quasar templates: A -- `Hot DOGs' \citep{tsai15}, B -- `featureless AGN' \citep{kirk12} and `obscuring torus' \citep{poll07}, C -- `type 2 quasar' \citep{poll07}, D -- `silicate AGN' \citep{kirk12} and E -- Mrk231 \citep{poll07}. Magenta curves with dots show the color locus of the linear combinations, for three different redshifts, of a star formation template with the HotDOG template. Dots mark 0, 10\%, 20\%, etc. contribution to the total 8-1000\micron\ infrared luminosity. All model colors include convolution of the templates with filter curves. Right: For three different redshifts (from top to bottom, $z=0.24$, 0.49 and 0.73 -- the bracketing redshifts of our sample, plus one value in the middle of the range), the relationship between the SF contribution to the apparent 160\micron\ flux as a function of its contribution to the bolometric luminosity. Even when star formation contributes only 20\% to the bolometric luminosity, over half of the apparent 160\micron\ flux is due to star formation for the redshifts of our sample.}
\label{pic_colors}
\end{figure*}

In Figure \ref{pic_colors}, left, we show infrared colors of our sources (black) compared with those of template star-forming galaxies (red) which are placed at the same redshift range as our targets ($z=0.24-0.73$). Quasars from our sample have noticeably warmer / bluer colors in the infrared than do star-forming galaxies. In principle, if we knew the spectral energy distribution of a `pure quasar' (i.e., a component that included the circumnuclear obscuring material where heating is dominated by the nucleus but excluded the larger host galaxy where heating is dominated by the stars) then from the observed colors of our objects we could determine the fractional contribution of the AGN and the host galaxy to each spectral energy distribution. 

To this end, we collect obscured AGN templates from the literature, including three from SWIRE \citep{poll07}: Mrk231 which has a power-law-like spectral energy distribution in the mid-infrared, `obscuring torus' which has a steep cutoff both at short and long wavelengths, and `type 2 quasar', which is obtained by heavy reddening of a type 1 quasar spectral energy distribution. These are supplemented by two more templates from \citet{kirk12}: `featureless AGN' that do not show silicate absorption, and `silicate AGN' which do. Finally, we also include the median spectral energy distribution of hot dust-obscured galaxies (HotDOGs) from \citet{tsai15}. The observations of these extremely luminous high-redshift obscured quasar candidates cover rest-frame wavelengths $\la 100$\micron, so in order to compute the 160\micron/70\micron\ colors we extrapolate the HotDOG template using a modified Rayleigh-Jeans spectrum with $\beta=1.5$ \citep{kirk12} beyond 100\micron. The model colors of AGN templates at $z=0.24-0.73$ are shown with blue lines in Figure \ref{pic_colors}. 

Intriguingly, half of the AGN templates (C -- `type 2 quasar', D -- `silicate AGN', and E -- Mrk231) are redder / colder than the observed colors of type 2 quasars. It is likely that a significant fraction of the total luminosity of these templates is due to the host galaxy instead of the active nucleus, so their colors are between those of the type 2 quasars in our sample and those of the star forming galaxies. `Featureless AGN' and `torus' templates (they have similar colors; marked B) and `HotDOGs' (marked A) have colors that are much closer to those observed in our sample. 

The magenta curves mark a linear combination of one of the star formation templates with the HotDOG template, going from 100\% of the 8--1000\micron\ luminosity dominated by star formation to 100\% dominated by the HotDOG template. The colors of our type 2 quasars are roughly consistent with such linear combinations if the quasar contributes at least half of $L_{\rm IR, 8-1000\mu m}$. Thus the observed infrared colors of type 2 quasars suggest that the bolometric luminosities of our objects are likely dominated by the quasar, not star formation in the host galaxy. 

However, because the spectral energy distribution of the quasar template declines so steeply beyond the peak, even a small fractional contribution of star formation is sufficient to dominate the observed 160\micron\ flux, as shown in Figure \ref{pic_colors}, right. As little as 20\% contribution of star formation to the total infrared luminosity $L_{\rm IR, 8-1000\mu m}$ is sufficient for it to contribute more than 50\% of the 160\micron\ flux at the redshifts of our sample. The lower the redshift, the longer is the rest wavelength probed by the 160\micron\ observations, and the smaller the contribution of star formation required to dominate at that wavelength. 

Unfortunately, these calculations do not allow us to unambiguously decompose the infrared spectral energy distribution of our sources into a quasar and star formation component, and to turn our upper limits on star formation rates into actual measurements of star formation rates. The primary reason is that the decomposition is sensitive to the assumed template for the quasar contribution, which is clear from the diversity of colors of AGN templates in Figure \ref{pic_colors}, left. A slight shift of the peak of the AGN template to longer wavelengths results in a larger contribution of the AGN to the 160\micron\ flux and to a smaller required contribution of star formation. The AGN templates are in turn sensitive to the geometry of the obscuring material (smooth vs clumpy, geometrically thin vs geometrically thick) and the relative orientation of the observer to the obscuring structure \citep{pier92, nenk02}. 

\citet{wyle15} conduct detailed spectral energy distribution decomposition of a subsample of 20 type 2 quasars from \citet{reye08} with {\it HST} observations most of which are also presented in this paper. The average and the standard deviation of the luminosities of the 20 objects are $L_{\rm [OIII]}=10^{43.1\pm 0.4}$ erg s$^{-1}$ and $\nu L_{\nu}$[12\micron]$=10^{45.0\pm 0.5}$ erg s$^{-1}$, so they represent the luminous end of the objects probed in this paper. Spectral energy distribution fits with CIGALE \citep{noll09} and DecompIR \citep{mull11b} suggest that in this subsample, the bolometric luminosities are dominated by the AGN (derived AGN fractions are $0.7\pm 0.2$ and $0.8\pm 0.15$ with the two methods, respectively). Nonetheless the median contribution of star formation to the observed 160\micron\ flux is 91\%, and the star formation rates derived by \citet{wyle15} are therefore similar to those we present here as upper limits. These conclusions are in agreement with our analysis based on far-infrared colors. While we continue treating our 160\micron-derived star formation rates as upper limits, we keep in mind that they are likely close to the actual star formation rates, even though the bolometric luminosities of our sources are dominated by the quasar. 

\subsection{Star formation rates of quasar hosts}

\begin{figure}
\centering
\includegraphics[scale=0.8, clip=true, trim=0cm 5cm 9.5cm 4cm]{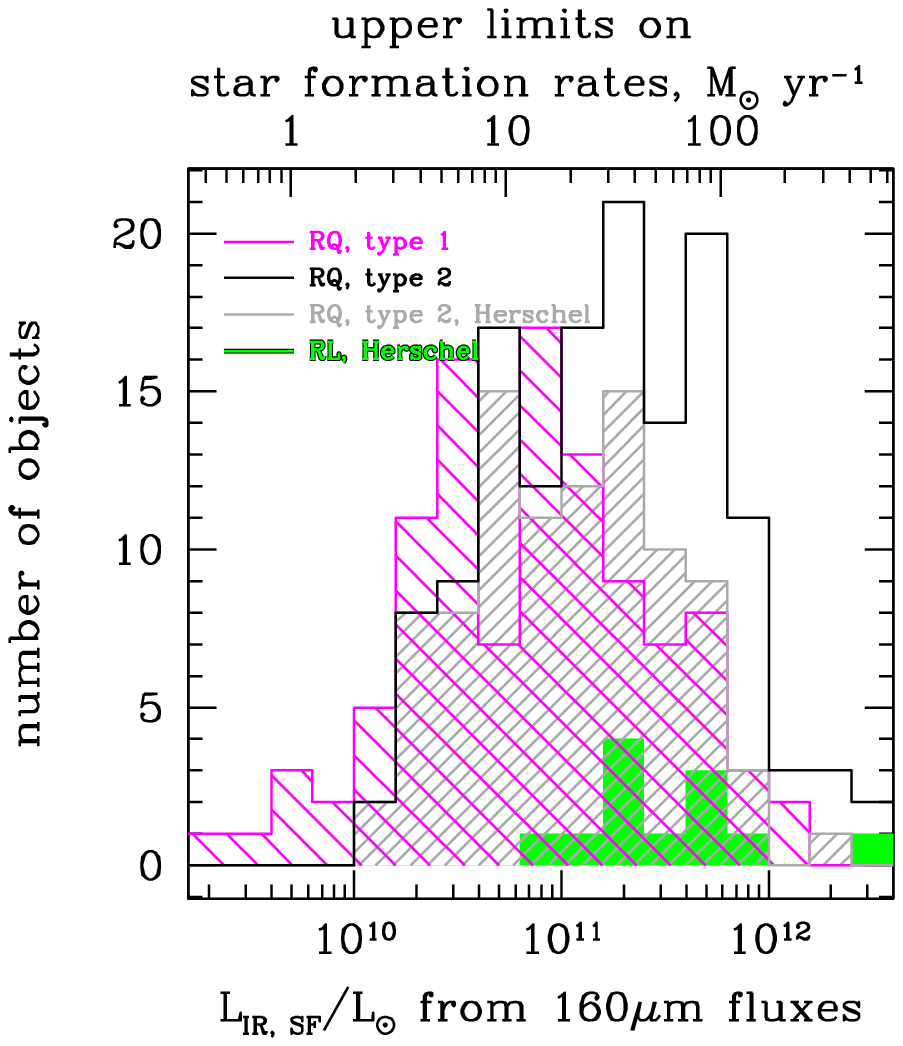}
\caption{Distributions of upper limits on infrared luminosities of star formation as derived from 160\micron\ fluxes. For radio-quiet type 1 quasars (magenta, sparsely shaded), the median (average) and the standard deviation are $\log (L_{\rm IR,SF,upper}/L_{\odot})=10.90(10.86)\pm 0.58$, corresponding to the median upper limit on star formation rate of 11.3$M_{\odot}$ yr$^{-1}$. Just for the blue type 1 quasars with deep \her\ observations \citep{petr15}, we find $\log (L_{\rm IR,SF,upper}/L_{\odot})=10.67(10.72)\pm 0.57$ and 6.3$M_{\odot}$ yr$^{-1}$. For radio-quiet type 2 quasars (solid black), $\log (L_{\rm IR,SF,upper}/L_{\odot})=11.24(11.22)\pm 0.54$. Excluding shallow \spi\ observations which include a lot of 160\micron\ non-detections and using only \her\ observations (grey, densely shaded), we find $\log (L_{\rm IR,SF,upper}/L_{\odot})=11.03(11.04)\pm 0.47$, corresponding to median upper limit on star formation rate of 18.1$M_{\odot}$ yr$^{-1}$. Type 1 and type 2 radio-loud sources (green, solid fill) with \her\ observations nominally show $\log (L_{\rm IR,SF,upper}/L_{\odot})=11.42(11.52)\pm 0.52$, but their 160\micron\ fluxes can be boosted by synchrotron emission and thus are unreliable measures of star formation.}
\label{pic_sfr}
\end{figure}

We compare the upper limits on star formation rates among the different subsamples of quasars discussed in this paper in Figure \ref{pic_sfr}. To convert from infrared luminosities of star formation to star formation rates, we use the calibration from \citet{bell03} which is slight modification of that of \citet{kenn98} and assumes Salpeter initial mass function. In this section we make a distinction between radio-quiet and radio-loud AGN by applying a simple luminosity cut $\nu L_{\nu}[1.4{\rm GHz}]=10^{41}$ erg s$^{-1}$ \citep{zaka04}. All sources (except one) with radio luminosities above this cutoff are detected in FIRST / NVSS, so we do not have to worry about upper limits on radio detections in potentially radio-loud sources. Of the 186 type 1 and type 2 quasars with \her\ data, twelve are radio-loud by this criterion. 

The first striking result is that the nominal star formation rates are higher in radio-quiet type 2 quasars than in radio-quiet type 1s. Part of this is due to the heterogeneity of our sample: a third of the type 2 quasar sample was observed with \spi-160, and these data have shallower observations and higher confusion limits than \her-160. As a result, 39 out of 51 type 2 quasars observed with \spi-160 are not detected. To make a better-matched comparison between the two samples, we consider only type 2 quasars observed with \her-160, the majority of which are detected. These objects still show appreciably higher star formation rates (or rather, upper limits on star formation rates) than type 1s with similarly deep \her\ observations and similar intrinsic luminosities and redshifts. Specifically, the median upper limits on star formation rate derived for hosts of blue type 1 quasars is 6$M_{\odot}$ yr$^{-1}$, whereas that of type 2 quasar hosts is 18$M_{\odot}$ yr$^{-1}$. (The nominal median upper limit on star formation in red type 1 quasars is even higher, 32$M_{\odot}$ yr$^{-1}$, but it is based on shallower MIPS-160 observations, with a third of the sample undetected.) As we discuss in Section \ref{sec:contrib}, even though our method technically only allows us to place an upper limit on the star formation rate, the actual values are likely close to the derived upper limits ($\sim 90\%$), so in the following discussion we make the assumption that these star formation upper limits are representative of the actual star formation rates. 

It is now well-established that star formation rates in obscured quasars are higher than those in unobscured ones \citep{kim06, zaka08, hine09, chen15}. This result appears to hold whether star formation is calculated from photometric or spectroscopic indicators. This conclusion is not well-explained by the classical orientation-based unification model, in which type 1 and type 2 quasars should occupy similar host galaxies. The difference in host star formation rates may be due to evolutionary effects: this observation could represent direct evidence that type 2 quasars are more likely to be found in dust-enshrouded environments characteristic of an ongoing starburst \citep{chen15}, as suggested by many models of galaxy formation \citep{sand88, hopk06}. 

An alternative explanation for a higher star formation rates in hosts of type 2 quasars is that the selection of these objects is biased toward gas-rich galaxies. It is usually assumed that AGN obscuration happens on circumnuclear scales ($\ll 1$ kpc), and it is not clear whether AGN obscuration is directly connected with the geometry of the host galaxy. In type 2 quasars, which occupy predominantly elliptical hosts, no clear relationship emerges between the presence or absence of the galactic disk and obscuration and their relative orientation \citep{zaka06, zaka08}. However, in less luminous type 2 AGN there are indications that at least some of the obscuration is occurring on the galactic scales by the gas and dust in the galaxy disk \citep{lacy07, lago11}. If this is a typical situation in type 2 quasars, then they would be preferentially found in more gas-rich and by extension more star-forming galaxies than type 1s. 

Finally, it is possible that type 1 and type 2 quasars discussed in this paper have different bolometric luminosities. While they have similar values of $\nu L_{\nu}$[12\micron], this measure could underestimate the luminosities of type 2 quasars which show evidence for obscuration even at mid-infrared wavelengths \citep{zaka08, liu13b}. Therefore, if the type 2s are significantly more luminous than the type 1s, and if the host star formation rate increases with quasar luminosity, then the observed difference in the host star formation rates may be due to the intrinsic differences between the two samples, but this scenario is unlikely given the very flat relationship between the AGN luminosity and the host star formation rates \citep{stan15}. Another possibility is that intrinsically more luminous type 2s (with apparent mid-infrared luminosities similar to those of type 1s) dominate their 160\micron\ fluxes, resulting in higher measured nominal star formation rates, but that is also unlikely in light of the spectral energy decomposition results by \citet{wyle15} who find that 91\% of 160\micron\ emission is due to star formation even in the luminous quasars as discussed in the previous section. 

Far-infrared star formation indicators are dominated by obscured star formation and may therefore miss unobscured star formation, which is normally estimated from the ultra-violet luminosities. Ultra-violet measures of star formation are extremely difficult to obtain in type 1 quasars, where the quasar in the nucleus of the galaxy makes identification of circumnuclear star formation all but impossible. Off-nuclear stellar populations, when measured with proper accounting for possible quasar light scattered off the interstellar medium into the observer's line of sight, tend to show post-starburst features \citep{cana13} rather than active star formation. In type 2 quasars, despite nuclear obscuration, ultra-violet emission is dominated by quasar scattered light \citep{zaka05, zaka06, obie15}. When this contribution is accounted for, the median ultra-violet rates of star formation in type 2 quasar hosts are $\la 3M_{\odot}$ yr$^{-1}$ \citep{obie15}, about an order of magnitude lower than those derived from far-infrared luminosities. This is in line with the typical balance between obscured and unobscured star formation found at $z<1$ \citep{burg13}. Therefore, we conclude that unobscured star formation is unlikely to affect our results.  

Another result apparent from Figure \ref{pic_sfr} is that the nominal upper limits on star formation rates of radio-loud quasars appear to be higher than those of radio-quiet quasars, which is contrary to many previous studies demonstrating that of all types of AGN, these objects are associated with the lowest rates of star formation \citep{shi07, dick12, dick14}. The reason for this discrepancy is that the 160\micron\ fluxes of these objects can be boosted by the synchrotron emission associated with the jet, not by the star formation in the host, particularly in type 1 quasars (four out of seven radio-loud objects) where jet emission might be expected to be beamed. As an example, we take a fiducial radio-loud quasar at $z=0.33$ with $\nu L_{\nu}$[1.4GHz]$=10^{43}$ erg s$^{-1}$ and spectral index of synchrotron emission of $\alpha=-0.5$ ($F_{\nu}\propto \nu^{\alpha}$). Such object would appear as a 2.3 Jy radio source at 1.4 GHz, and if its synchrotron spectrum continues all the way to the wavelengths of our far-infrared observations then its 160\micron\ flux due to the jet would be 61 mJy. With our one-band observations, we would calculate a star formation rate of 41 $M_{\odot}$ yr$^{-1}$. The exact contribution of synchrotron emission to the 160\micron\ flux depends sensitively on the spectral index and on whether the synchrotron spectrum continues all the way to these high frequencies. We conclude that upper limits on star formation rates derived from 160\micron\ data are not very meaningful for radio-loud objects; their star formation rates are likely lower than those reported in Figure \ref{pic_sfr}.  

\section{Star formation rates of quasar hosts from spectroscopy}
\label{sec:spec_analysis}

In this section we continue investigating star formation rates in quasar host galaxies (and the resulting radio emission), but now using mid-infrared spectroscopic data. We use polycyclic aromatic hydrocarbon features and low-ionization emission lines as potential star formation diagnostics. 

\subsection{Calibration of PAHs as star formation diagnostics}

Polycyclic aromatic hydrocarbon (PAH) emission dominates mid-infrared spectra of star-forming and starburst galaxies and is typically seen as a star-formation indicator \citep{alla89, genz98, rous01, peet04, kenn09}. Various studies have presented calibrations between PAH luminosities and infrared luminosities of star formation or star formation rates \citep{bran06, farr07, shi07, hern09, felt13}, but often either the PAH luminosities or the infrared luminosities or the star formation rates are not measured exactly the same way as what we do here, resulting in significant systematic uncertainties. Most notably, some authors use broad-band 8\micron\ fluxes \citep{rous01, kenn09} which likely overestimate the amount of PAH[7.7\micron] emission. Others (e.g., \citealt{bran06, hern09}) measure PAH luminosities spectroscopically by subtracting an interpolated continuum, which likely results in an underestimate of the PAH fluxes since Drude-like PAH profiles \citep{smit07} have significant flux in the wings. For example, for the 6.2\micron\ feature in ultraluminous infrared galaxies \citep{hill14} we calculate the PAH flux using the Drude-fitting method, as well as by subtracting a linear continuum interpolated between 5.95\micron\ and 6.55\micron, and we find that the former is a factor of 2 higher than the latter. 

Of the PAH / star formation calibrations presented in the literature, the one that utilizes the measures closest to ours is presented by \citet{shi07}. These authors calculate PAH luminosities by fitting Drude profiles to the PAH features, similarly to our procedure, and they find an appropriate star formation template with the same PAH luminosity to calculate the star formation luminosity between 8 and 1000 \micron. Their final conversion between PAH[11.3\micron] and $L_{\rm IR, SF}$ is well described by 
\begin{equation}
\log({\rm PAH[11.3\micron]}/L_{\odot})=-0.7842+0.8759\log(L_{\rm IR,SF}/L_{\odot}).\label{eq_shi}
\end{equation}
This calibration can then be supplemented by the IR-to-radio correlation from \citet{bell03} listed in equation (\ref{eq_ir_radio}) to obtain a direct relationship between PAH luminosities and radio emission due to star formation:
\begin{equation}
\log L_{\rm radio,SF}{\rm [erg\, s^{-1}]}=27.4584+1.2620 \log ({\rm PAH[11.3\micron]}/L_{\odot}).\label{eq_pah}
\end{equation}
The infrared luminosities due to star formation can be converted to star formation rates \citep{bell03}. 

To double-check calibrations (\ref{eq_shi})-(\ref{eq_pah}), we use nearby star-forming galaxies from \citet{bran06} who tabulated $L_{\rm IR, SF}$ values measured from the Infrared Astronomical Satellite (IRAS; \citealt{neug84}) data. Of the 22 objects presented in their paper, we exclude four with the optical signature of an active nucleus. For the remaining 18, we download the reduced spectra from the Spitzer-ATLAS database \citep{hern11}, convolve them with IRAS-12\micron\ and IRAS-25\micron\ filter curves from \citet{dopi05} to calculate synthetic IRAS fluxes and compare them with those observed. We then augment the spectra by a factor necessary to match the synthetic fluxes to the observed fluxes and use the resulting flux-calibrated spectra to measure PAH luminosities using our Drude-fitting methods. We also obtain the highest listed 1.4 GHz radio fluxes for these galaxies from the NASA/IPAC Extragalactic Database. We assume that any major flux discrepancies between the listed 1.4 GHz measurements are due to varying resolutions of the radio data; for these nearby sources, we prefer the lowest resolution, highest flux observations which are more likely to capture extended radio flux. 

On the high-luminosity end, we use a subsample of GOALS galaxies from \citet{stie14} dominated by star formation (EW of PAH[6.2\micron] $>0.3$\micron) and hosted by single or late-stage merger hosts \citep{stie13}, as was described in Section \ref{sec:ir_radio}. We further restrict our analysis to objects with silicate absorption strength $<1.1$, to avoid any biases in PAH measurements due to our approach of not correcting PAH fluxes for extinction. Although \citet{stie13} present PAH flux measurements, we have found that even minor differences between their and our fitting procedures result in a noticeable systematic offset in PAH/$L_{\rm IR,SF}$ and PAH ratios. Therefore, we re-measure PAH features in these objects using exactly the same tools as those used for type 1 and type 2 quasars. We download the IRS enhanced data products from the \spi\ Heritage Archive, stitch together the SL and LL orders, flux-calibrate using IRAS-12\micron\ and IRAS-25\micron\ fluxes and measure PAHs using our Drude-fitting methods. The radio fluxes of these sources are available in \citet{u12} and in NVSS. Flux calibration of \citet{bran06} and \citet{stie14} spectra against IRAS data is necessary only because some of these objects are so nearby that they are extended beyond the IRS slits. For type 2 quasar spectra, most of which appear as point sources to \spi, the default flux calibration provided for the enhanced data products of IRS is in excellent agreement with \spi\ and WISE photometry, and the flux calibration step is unnecessary. 

\begin{figure}
\centering
\includegraphics[scale=0.45, clip=true, trim=0cm 10cm 0cm 0cm]{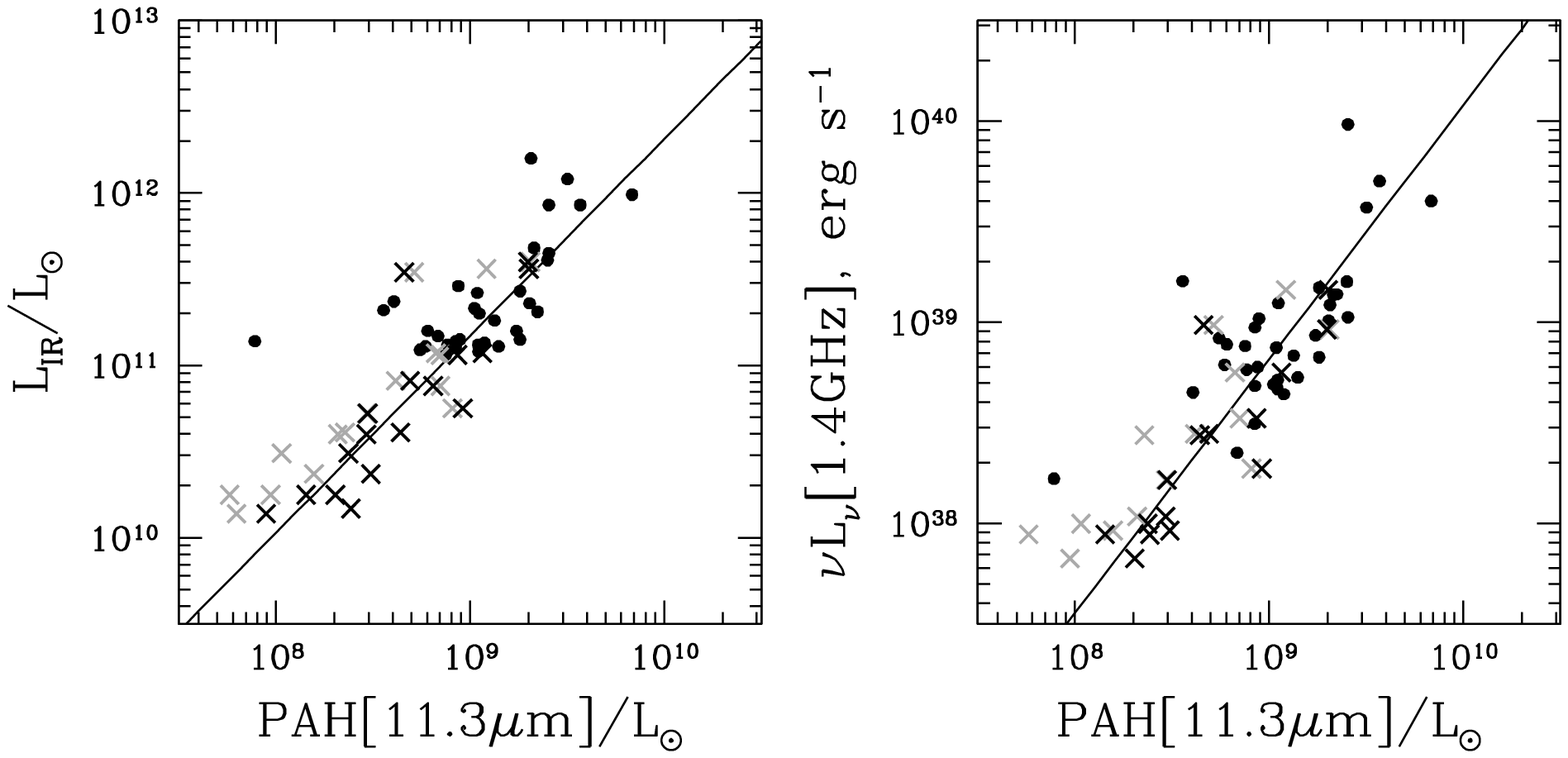}
\caption{Double-checking the PAH vs infrared and radio correlations for star-forming galaxies. The solid lines show our adopted calibrations (\ref{eq_shi}) and (\ref{eq_pah}). Crosses show star-forming galaxies from \citet{bran06}, in black for results obtained using IRAS-12 for bolometric flux calibration and in grey using IRAS-25. Circles show the star-forming subsample of the GOALS galaxies from \citet{stie14}. Star-forming galaxies of low and high luminosities show a good agreement with our adopted calibrations, with standard deviations $\la 0.3$ dex. }
\label{pic_pah_calib}
\end{figure}

As seen in Figure \ref{pic_pah_calib}, we find good agreement between the scaling relationships (\ref{eq_shi}) and (\ref{eq_pah}) and the actual measurements in these two samples of star-forming galaxies. Given a measurement of PAH[11.3\micron], our adopted calibrations (\ref{eq_shi}) and (\ref{eq_pah}) predict the total infrared and radio luminosities for star-forming galaxies with a standard deviations of $\la 0.3$ dex. Thus the PAH vs star formation conversion appears to have somewhat higher intrinsic spread than the conversion between 160\micron\ flux and $L_{\rm IR,SF}$ and $L_{\rm radio, SF}$, which has a standard deviation of $\la 0.2$ dex. Much of this spread likely reflects the intrinsic dispersion of galaxy properties, as the PAH features are strongly detected in all sources, the quality of the data are high, and PAH measurement uncertainties are only a few per cent for this sample.

\subsection{PAH measurements of star formation in quasar hosts}

\begin{figure*}
\centering
\includegraphics[scale=0.7, clip=true, trim=0cm 10cm 10cm 0cm]{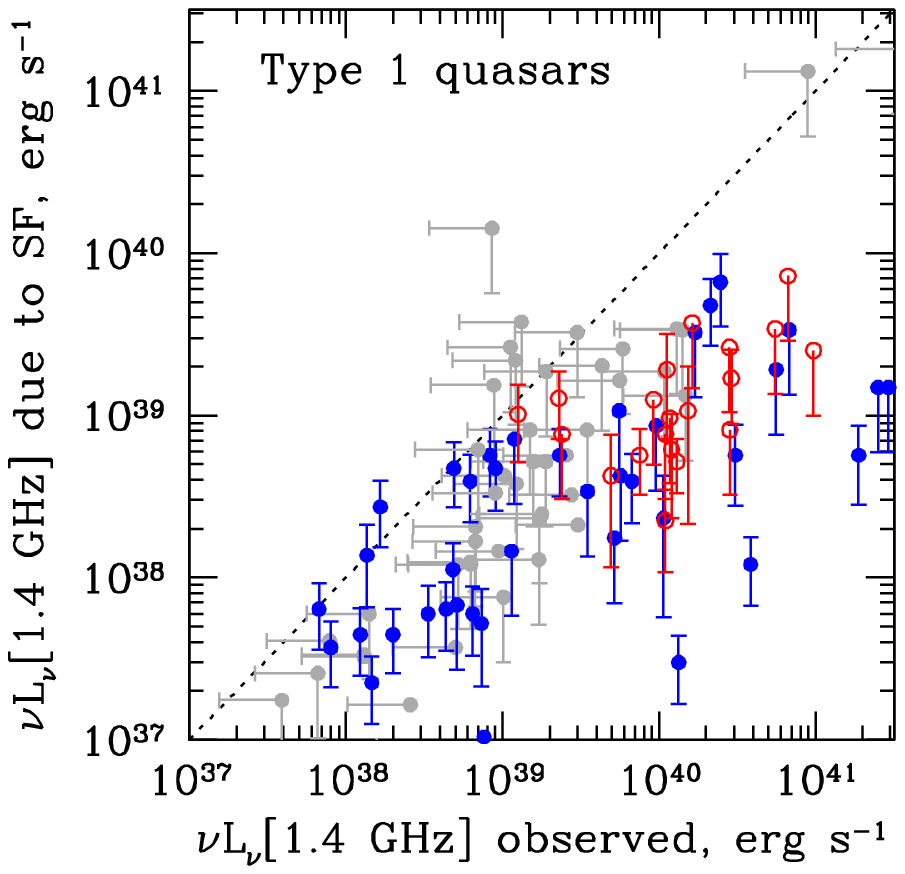}%
\includegraphics[scale=0.7, clip=true, trim=0cm 10cm 10cm 0cm]{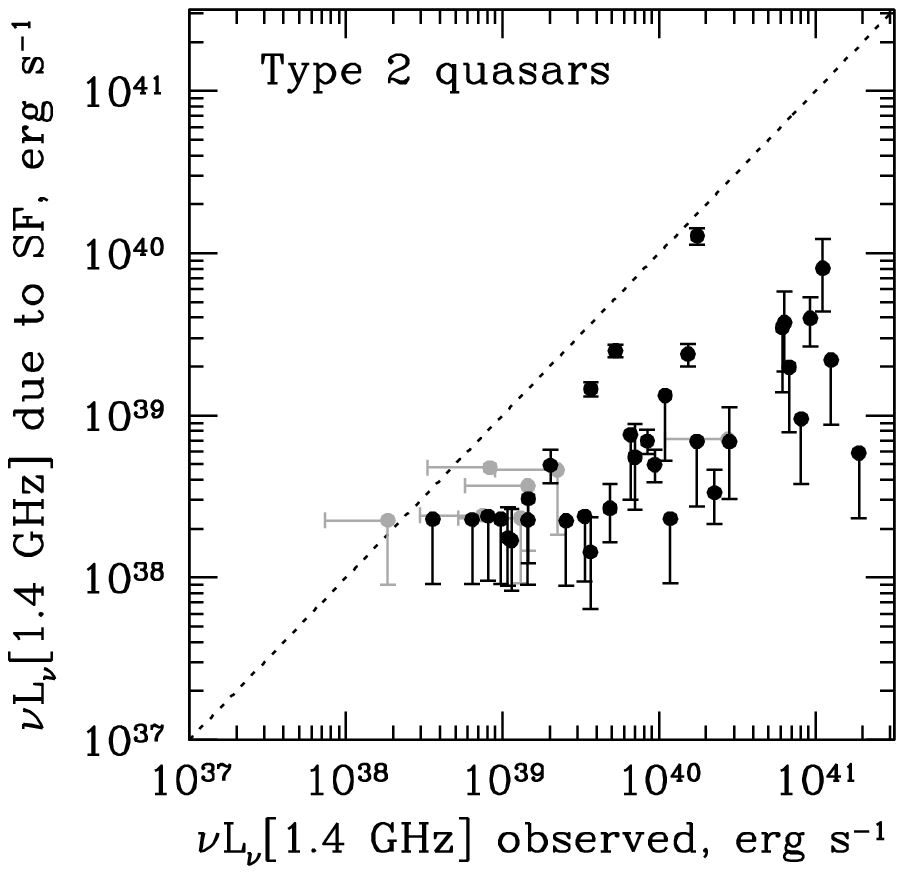}\\
\caption{Results from PAH-based measures of star formation. Left: Radio luminosities expected from star formation and the observed radio luminosities for PG (solid blue circles) and 2MASS (open red circles) quasars from the sample of \citet{shi07}. Objects not detected in the radio are shown in grey. Right: same for the type 2 quasars from \citet{reye08}. The majority of all quasars lie below the 1:1 dashed line, indicating that star formation in quasar host galaxies is insufficient to account for the observed radio emission from quasars, by 1.1-1.3 dex on average.}
\label{pic_shi_pah}
\end{figure*}

As the sources in our sample are quasars with $L_{\rm bol}\ga 10^{45}$ erg s$^{-1}$, their continuum emission in the mid-infrared is dominated by the thermal emission of quasar-heated dust, with PAH features sometimes visible on top. In this Section, we start by using the 11.3\micron\ PAH feature exclusively and in Section \ref{sec_issues} we discuss the reliability of this measure. In Figure \ref{pic_shi_pah}, left, we show the predicted radio luminosity due to star formation for type 1 quasars analyzed by \citet{shi07}, and in the right panel we show the same for type 2 quasars. 

With the exception of a handful of problematic objects, our task for the \citet{shi07} sample is straightforward: we take their measured PAH-derived total luminosities of star formation, convert them to the expected radio luminosity of star formation (eq. \ref{eq_ir_radio}) and compare with the observed values. We use most of their $L_{\rm IR, SF}$ values, with the following exceptions. We add an upper limit to star formation for PG 0003+158, where we use their upper limit on PAH[11.3\micron]. Furthermore, for 2MASS J130700.66+233805.0 and 2MASS J145331.51+135358.7 we do not use their PAH[7.7\micron] measurements which are overestimated due to unmodeled ice absorption and instead use their PAH[11.3\micron] measurements; this results in a decrease of the calculated $L_{\rm IR,SF}$ for the former object. For type 2 quasars, we use our own PAH[11.3\micron] measurements and convert them to the expected radio emission using equation (\ref{eq_pah}). 

Figure \ref{pic_shi_pah} makes it clear that if PAH[11.3\micron] is a good measure of the host galaxies' star formation rates, then in all quasar samples (blue type 1s, red 2MASS-selected quasars and obscured type 2s) the observed radio emission is well in excess of that expected from the star formation, which is consistent with the results we obtained from photometric measures of star formation in Section \ref{sec:rem_photo}. The median / mean / sample standard deviation ratio of the observed radio emission to that predicted from star formation is 1.1 dex / 1.2 dex $\pm$ 0.9 dex for quasars in the left panel and 1.1 dex / 1.3 dex $\pm$ 1.1 dex for quasars in the right panel (only objects with radio detections were taken into account). 

PAHs are detected in half of each subsample -- blue type 1s, red type 1s and type 2s. Among the detections, the median star formation rates follow the trend seen in photometric data (lower star formation rates in the blue type 1 subsample than in the other two): 6.7, 26 and 29 $M_{\odot}$ yr$^{-1}$, correspondingly. Based on the same dataset, it was pointed out by \citet{shi07} that red type 1 quasars occupy more actively star-forming hosts than blue type 1s. We convert non-detections to upper limits on star formation rates, but in some cases the quality of the data make them not very constraining, and the median upper limits are 35, 64 and 17 $M_{\odot}$ yr$^{-1}$. 

\subsection{Issues with PAH measures of star formation}
\label{sec_issues}

\subsubsection{PAHs and dust obscuration}

It is not clear to what extent PAH emission is affected by intervening dust absorption. \citet{zaka10} showed that in star-forming ultraluminous infrared galaxies, ratios of PAH complexes at different wavelengths are correlated with the strength of the silicate absorption feature, in a manner consistent with PAH-emitting regions being affected by an amount of obscuration similar to (though somewhat smaller than) that affecting the mid-infrared continuum; this possibility was also pointed out by \citet{bran06} and \citet{iman07}. But this trend is not borne out in a large sample of lower-luminosity GOALS galaxies \citep{stie14}. Perhaps only the most luminous, most compact starbursts follow a relatively simple geometry in which both the PAH-emitting regions and the thermal-continuum-emitting regions are enshrouded in similar amounts of cold absorbing gas, whereas in more modest star-forming environments the different components are mixed with one another. Therefore, it remains unclear whether PAH emission is affected by absorption, and any absorption correction would be rather uncertain. 

Because type 2 quasars show weak silicate absorption \citep{zaka08}, any putative absorption corrections to PAH fluxes are relatively minor: in 90\% of our sample, the peak absorption strength is $S[9.7\micron]<1$, so the optical depth at 11.3\micron\ is less than 0.6 and the correction to PAH flux would be less than 30\%, with a median of 8\%. But more importantly, in type 2 quasars most of the mid-infrared continuum is due to the quasar-heated dust, and therefore the strength of silicate absorption reflects the geometry of this circumnuclear component. Correcting PAH emission which is likely produced in different spatial regions using this value of silicate absorption appears meaningless, so we do not attempt it. 

If star formation in quasar hosts is extremely obscured and if PAH-emitting regions are buried inside optically thick layers of dust, then our methods would underestimate PAH emission, and thus the star formation rate and the expected radio emission. In order for such absorption to completely account for a 1.2 dex offset between the observed and the predicted radio emission, we would need an optical depth of 2.8 at the wavelength of the 11.3\micron\ feature, which corresponds to the strength of the silicate absorption of about $S[9.7\micron]\simeq 4$. Only 6\% of ultraluminous infrared galaxies have an apparent absorption strength above 3 \citep{zaka10}, so such high average values of extinction toward PAH-emitting regions are implausible. Therefore, it is unlikely that using the PAH method we underestimate the star formation rates in quasar hosts by the amount necessary to explain the difference between the observed and predicted radio emission. 

\subsubsection{Effect of the quasar radiation field on PAHs}
\label{sec:effect}

\begin{figure}
\centering
\includegraphics[scale=0.45, clip=true, trim=0cm 10cm 0cm 0cm]{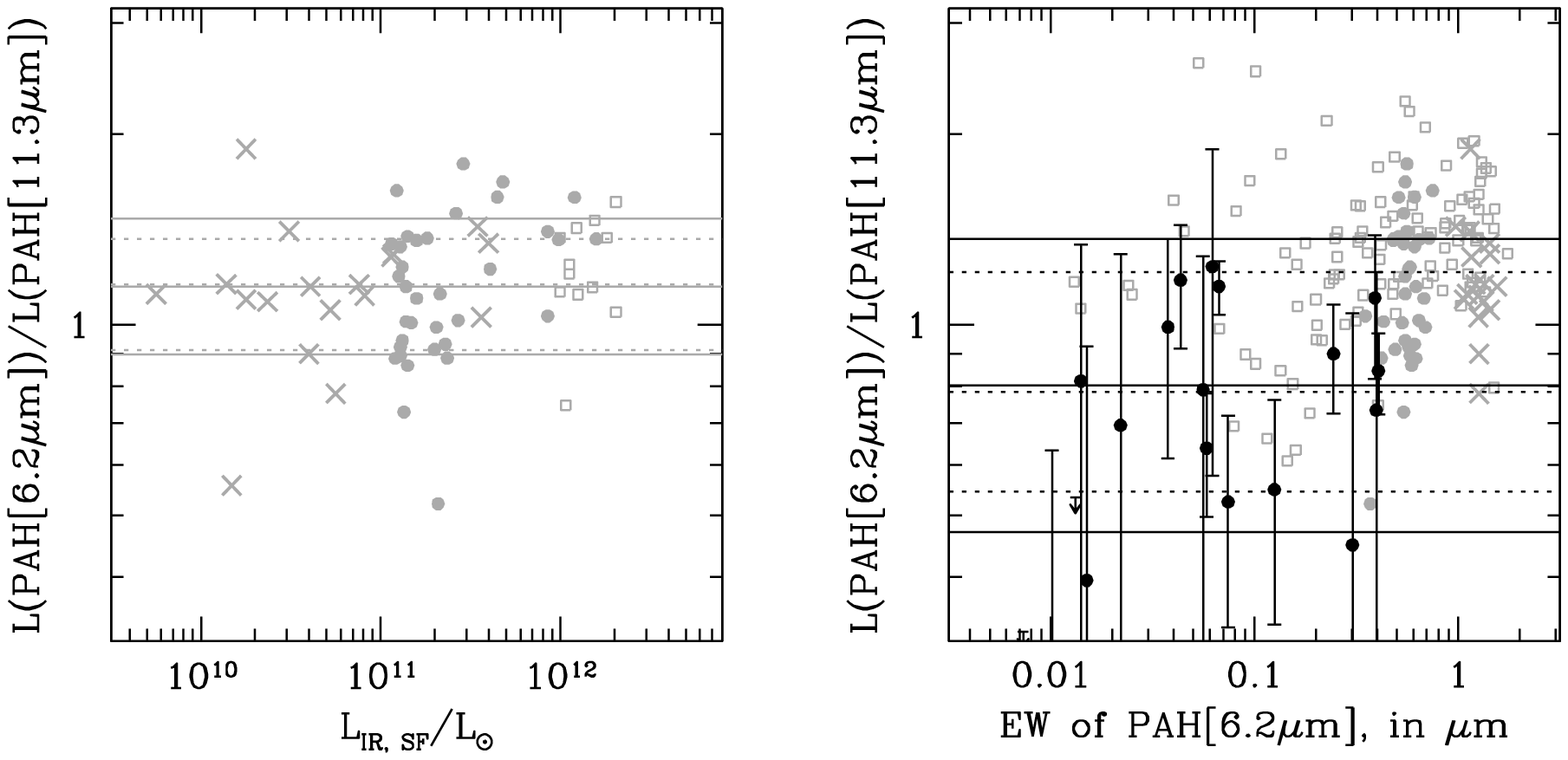}
\caption{Left: PAH[6.2\micron]/PAH[11.3\micron] ratios in star-forming galaxies from our comparison samples (crosses: \citealt{bran06}, circles: \citealt{stie14}, squares: \citealt{hill14}) and the median and 68\% range derived from these data (solid horizontal lines) and those presented by \citet{diam10}, shown in dotted horizontal lines. Right: PAH ratios in the comparison samples (\citealt{bran06, stie14, hill14}, grey), whether or not they have any signs of an AGN, and those of type 2 quasars in our spectroscopic sample (black points with error bars). The horizontal lines denote the mean and the standard deviation for type 2 quasars (solid lines) and the same for Seyfert galaxies (dotted lines) as presented by \citet{diam10}. PAH ratios of type 2 quasars are more in line with those in Seyferts than those in star-forming galaxies, although there is not a strong correlation between the PAH ratios and the relative contribution of the active nucleus.}
\label{pic_pah_ratios}
\end{figure}

The harsh radiation field of the quasar and shocks due to quasar-driven outflows may destroy PAH-emitting molecules and therefore suppress PAH emission \citep{smit07, diam10, lama12}. As a result, by using PAH luminosities we may be underestimating the star formation rates in quasar hosts. To evaluate this possibility, in Figure \ref{pic_pah_ratios} we investigate the ratios of PAH[6.2\micron] to PAH[11.3\micron] in star-forming galaxies and in quasars. Because these ratios are related to the size distribution of the aromatic molecules, differences between PAH ratios found in star-forming galaxies and those found in quasars may indicate that the quasar has an impact on the PAH-emitting particles and make PAH-based star formation rates suspect. 

In the left panel of Figure \ref{pic_pah_ratios}, we demonstrate that our measurements of the median and the standard deviation of the PAH[6.2\micron]/PAH[11.3\micron] ratio among star forming galaxies are in excellent agreement with those of \citet{diam10}, whose PAH-fitting procedures are very similar to ours. For this comparison, we pre-selected objects without any optical or infrared signatures of an AGN, so that we are evaluating purely star-forming galaxies. In the right panel, we now include PAH measurements of sources with varying degree of AGN contribution, and demonstrate the dependence of the PAH ratios as a function of the EW of PAH[6.2\micron], a common measure of the fractional AGN contribution to the bolometric budget \citep{spoo07}. Because AGN tend to produce power-law-like emission in the mid-infrared, while star-forming galaxies (which have lower dust temperatures) rarely have strong continuum at these wavelengths, the EW of PAH[6.2\micron] is high in star-forming galaxies and low in AGN, with 0.3\micron\ being the typical dividing line \citep{petr11, stie13, stie14}. 

We see that type 2 quasars display relatively low EW of PAH[6.2\micron], in agreement with our previous conclusion that their bolometric luminosities are dominated by AGN activity. Furthermore, despite large measurement errors, we find that the PAH[6.2\micron]/PAH[11.3\micron] ratio is suppressed in type 2 quasars, and again we see excellent quantitative agreement between our measurements of the median ratio and its standard deviation and those of \citet{diam10}. The relatively low PAH[6.2\micron]/PAH[11.3\micron] in type 2 quasars cannot be explained by extinction which would increase this ratio: due to the silicate feature centered at 9.7\micron\ which extends over a wide wavelength range, dust opacity is higher at 11.3\micron\ than at 6.2\micron\ \citep{wein01, chia06}. Therefore, we confirm that the PAH ratios appear to be affected by the quasar radiation field. However, it is not clear how much of an effect quasar radiation field may have specifically on the PAH[11.3\micron]-derived star formation rates. \citet{diam10} argue that PAH[11.3\micron] feature may be less affected than PAH[6.2\micron] which traces smaller easily destroyed grains, but \citet{lama12} have argued that PAH[11.3\micron], too, can be suppressed by the quasar emission resulting in an underestimate of the star formation rate. 

\begin{figure}
\centering
\includegraphics[scale=0.45, clip=true, trim=0cm 10cm 10cm 0cm]{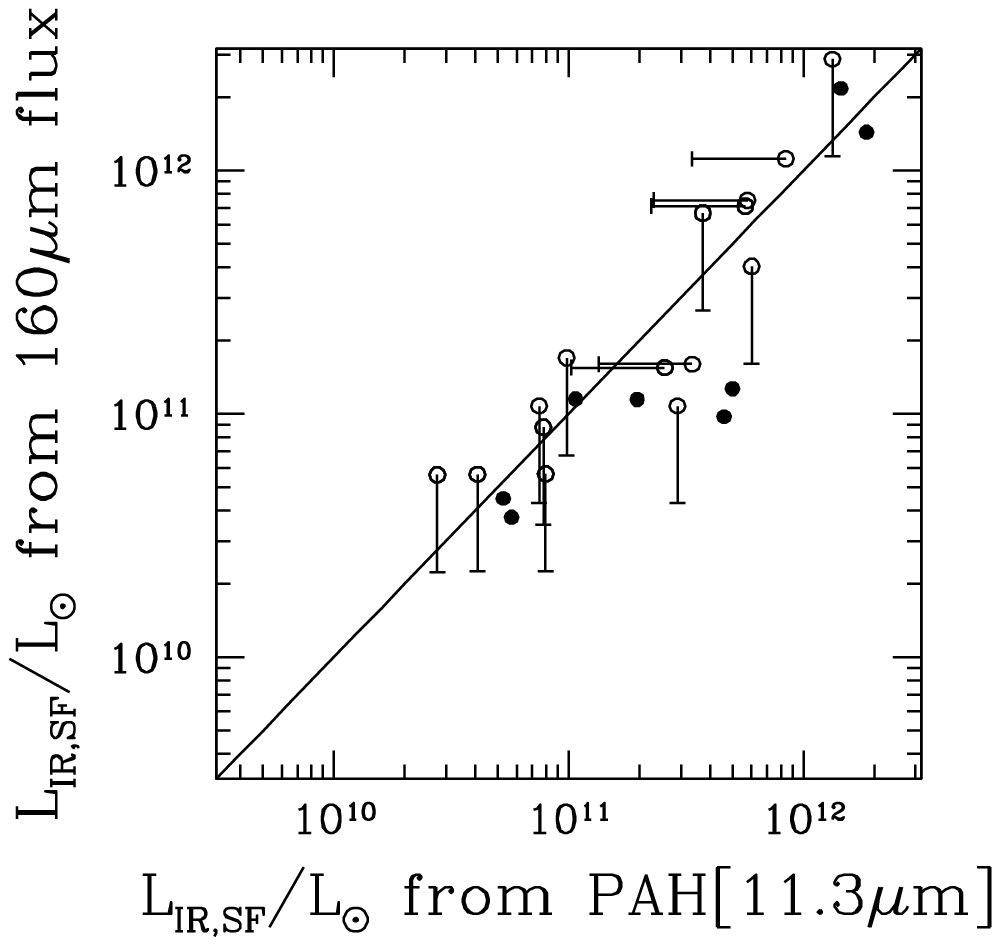}
\caption{Following \citet{petr15}, we directly compare infrared luminosities of star formation in type 2 quasar hosts as derived from PAH[11.3\micron] and from the 160\micron\ flux. Solid circles show type 2 quasars with both a 160\micron\ detection and a PAH[11.3\micron] detection, whereas open circles denote upper limits in one of these measures (those objects with upper limits in both PAH[11.3\micron] and 160\micron\ flux are excluded). The solid line shows the locus expected for star forming galaxies. In type 1 quasars, \citet{petr15} find that 160\micron\ photometry systematically overestimates star formation rates, or PAH[11.3\micron] measurements systematically underestimate them, or both. This is not borne out by the data for type 2 quasars, although the sample is too small for a conclusive investigation.}
\label{pic_flux_irs}
\end{figure}

This is suggested by \citet{petr15} who show that the star formation rates of type 1 quasar hosts derived from far-infrared luminosities are a factor of two-three higher than those derived from PAH[11.3\micron] luminosities. Thus either the far-infrared fluxes are contaminated by the quasar resulting in an overestimate by this factor, or the PAH-emitting particles are destroyed, resulting in an underestimate by this factor, or perhaps both are true to a lesser extent. In our sample of type 2 quasars, there is unfortunately limited overlap between objects with 160\micron\ photometry and IRS spectroscopy (28 objects, most of them with upper limits in at least one of these measurements). In the few that do have both measures of star formation (Figure \ref{pic_flux_irs}), we do not detect any noticeable offset from the locus of star-forming galaxies, suggesting that both 160\micron\ fluxes and PAH[11.3\micron] luminosities provide consistent measures of star formation rates in these objects. 

\subsection{Star formation rates from [NeII] and [NeIII] lines}
\label{sec:neon}

\begin{figure}
\centering
\includegraphics[scale=0.45, clip=true, trim=0cm 10cm 0cm 0cm]{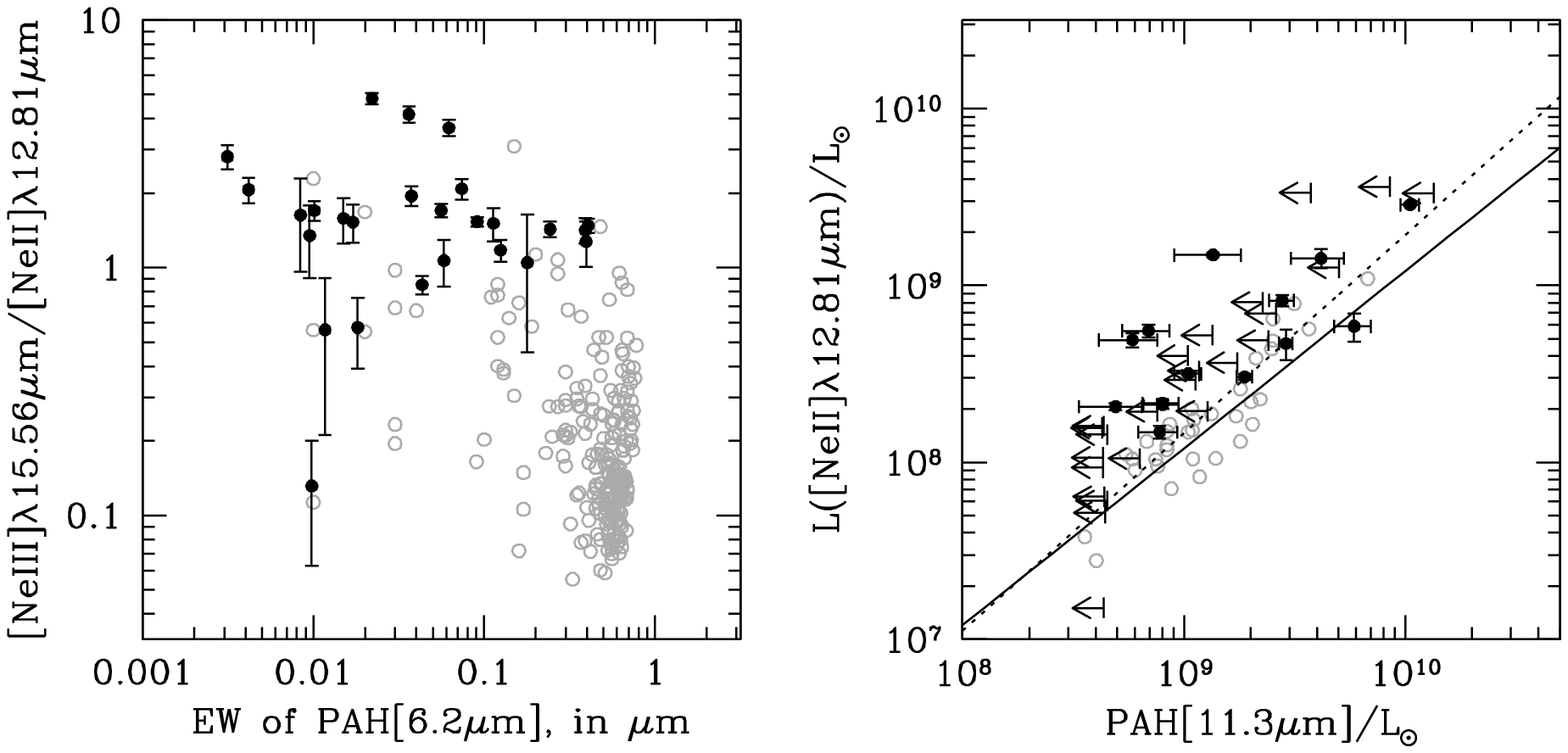}
\caption{Left: Neon ratios in type 2 quasars (solid black points) and in GOALS galaxies (open grey points, \citealt{inam13}), as a function of the equivalent width of the PAH[6.2\micron] feature. Objects with EW of PAH[6.2\micron] of $<0.3\micron$ (which includes the majority of type 2 quasars) are thought to be bolometrically dominated by the AGN. We find that these sources tend to show significantly higher [NeIII]/[NeII] ratios than those seen in pure star-forming galaxies, suggesting that AGN contributes to neon ionization. Right: Two mid-infrared star-formation diagnostics, PAH[11.3\micron] luminosity and [NeII] luminosity, plotted against each other for star-forming galaxies from the GOALS sample (open grey points) and for type 2 quasars (solid black points and arrows), along with two star formation calibrations provided in the literature: the solid line is from \citet{diam10} and the dotted line is from \citet{ho07}. GOALS star forming galaxies show excellent agreement with both calibrations, but type 2 quasars show either an excess of [NeII] or a deficit of PAH[11.3\micron] by about 0.35 dex, or a factor of 2.2. We choose to compare [NeII] (which is an over-estimate of star formation) with PAH[11.3\micron] (which may be an underestimate of star formation) rather than 160\micron\ fluxes (which provide upper limits on star formation) because these two values bracket actual star formation rates.}
\label{pic_ne_ratios}
\end{figure}

Low-ionization forbidden emission lines are frequently used as a star-formation diagnostic. In Figure \ref{pic_ne_ratios} we examine the utility of [NeII]$\lambda$12.81\micron\ and [NeIII]$\lambda$15.56\micron\ for use as star formation diagnostics in our sample. The line ratios of neon are clearly affected by the presence of the quasar: [NeIII]/[NeII] ratios are significantly higher in our sample of type 2 quasars than in the comparison star-forming galaxies (left). Furthermore, [NeV]$\lambda$14.32\micron\ and [NeVI]$\lambda$7.65\micron\ are frequently strongly detected in our sources, with [NeV]/[NeII]$\sim 1$, whereas these lines are rarely seen at all in star-forming galaxies \citep{farr07}. 

Thus [NeIII] and higher ionization lines are clearly affected by the presence of the quasar. In Figure \ref{pic_ne_ratios}, right, we explore whether [NeII] can still be safely used as a star formation diagnostic or whether it, too, has an appreciable contribution from the quasar. Using GOALS star-forming galaxies, we reproduce with high accuracy the PAH vs [NeII] calibrations from the literature \citep{ho07, diam10}. The median offset between GOALS galaxies and these relationships is only $\sim 0.04$ dex, with a standard deviation around the relationships of $\sim 0.17$ dex, i.e., the quality of the PAH-[NeII] correlation is comparable to, or better, than the correlations between other star formation indicators we have discussed here.

In contrast, type 2 quasars lie systematically above the star formation relations, with a median offset of about 0.3 dex, or a factor of 2. Therefore, either the quasar contributes appreciably to the [NeII] ionization, or PAH[11.3\micron] features are suppressed, or both. Because we have not found any offset between PAH-derived and 160\micron-derived star formation rates (Figure \ref{pic_flux_irs}), we suggest that the destruction of PAH[11.3\micron] may not be the dominant effect and that instead the quasar dominates the photoionization of [NeII]. This finding is similar to that of \citet{kim06} who found that low-ionization optical emission lines, e.g., [OII]$\lambda$3727\AA, may also be dominated by quasar photoionization and be poor star formation indicators. 

However, the possible contribution of the quasar photoionization to the [NeII] emission does not modify our main conclusions regarding the excess of radio emission over that predicted due to star formation. Even if we were to use [NeII] as a star formation diagnostic, we would still underpredict the amount of radio emission in type 2 quasars by 0.42 / 0.78 dex (median / mean), or a factor of 2.6 / 6.0. 

\section{Discussion: effects of quasar on star formation and its diagnostics}
\label{sec:discussion}

Although spectral energy distribution decomposition methods assume that an AGN component and a star-forming component are simply added together to produce the overall galaxy spectrum, the two components could affect each other in a much less linear way. For example, a powerful but compact circumnuclear starburst could be further heated by the AGN, raising the apparent dust temperature beyond those encountered in star-formation templates and leading the observer to classify the source as quasar-dominated. A dusty galaxy with modest star formation rates could host a quasar; with a fortuitous geometric distribution of dust, the quasar can be hidden from view and lead to strong far-infrared emission from dust on $\ga$ kpc scales, which can be mistaken for high rates of star formation. PAH-emitting particles could be destroyed by the harsh radiation from the AGN, while low-ionization emission line regions associated with star formation can get photo-ionized by the quasar, suppressing these star-formation diagnostics. 

We find clear evidence for some of these effects: quasars likely suppress some PAH emission, enhance the low-ionization emission-line diagnostics, and contribute to the far-infrared emission, thereby likely affecting all mid- and far-infrared diagnostics of star formation. To minimize these effects, we choose the longest wavelength observations available to us (160\micron) which have the smallest fractional contamination by the AGN, PAH[11.3\micron] which is less affected by the AGN than PAH[6.2\micron] and less contaminated by high-ionization line emission than PAH[7.7\micron], and [NeII]$\lambda$12.81\micron. These star formation measures agree with one another to within a factor of two even in quasar hosts (with [NeII] likely the most strongly affected by the quasar), giving some credence to our measured rates of star formation. Therefore, we find it highly unlikely that we have underestimated star formation rates in quasar hosts by a factor of 10, the value required to bring the observed radio fluxes in agreement with those expected from star formation alone. 

The impact of quasars on the evolution of their host galaxies has emerged as a key question in modern galaxy formation models. Active black holes are now suspected in limiting galaxy masses via quasar feedback \citep{tabo93, silk98, spri05}, and determining the progression of star formation activity during and after an episode of black hole activity remains an interesting and unsolved problem in observational astronomy. Some studies suggest that long-term average AGN luminosity and star formation are strongly correlated, potentially due to a common supply of cold gas \citep{raff11, mull12, chen13}. In individual AGN, the correlations between AGN luminosity and their hosts' star formation rates are quite weak \citep{shao10, rosa12, harr12b, stan15}, which can largely be explained by short-term fluctuations in AGN luminosity \citep{hick14}. Intriguingly, some studies have reported an apparent suppression of star formation in quasar hosts \citep{page12, barg15}, although in some cases this can be attributed to limited survey volumes and sample sizes \citep{harr12b}. In any case, a potential detection of a suppression in star formation for quasar hosts must be distinguished from over-ionization or destruction of the star-formation diagnostics by the quasar. This is necessary in order to be able to measure the impact of the quasar on its host galaxy. 

Quasars with luminosities $\ga 10^{46}$ erg s$^{-1}$ are luminous enough to easily ionize the gas over galaxy-wide scales \citep{liu09,gree11,hain13,hain14}, and quasar-driven outflows \citep{liu13a,liu13b} can lead to galaxy-wide shocks \citep{rich11,hill14, zaka14}. While circumnuclear obscuration could shield parts of the galaxy from direct quasar emission, outflows and shocks might find ways around this obstacle \citep{wagn13}. Physical removal of the interstellar medium suppresses subsequent star formation activity, but it would also be interesting to detect the effect of the quasar on on-going star formation in the regions affected by quasar radiation and / or quasar-driven winds. One possibility is that star formation proceeds as usual, but emission line diagnostics of star formation are strongly affected by quasar photoionization or shock excitation. This could be the origin of the decreasing [OII]$\lambda$3728\AA/[OIII]$\lambda$5007\AA\ ratio both as a function of quasar luminosity \citep{ho05, kim06} and narrow-line kinematics \citep{zaka14}. If this hypothesis is correct, then one can hide significant amounts of star formation by exposing these regions to the quasar radiation field which would bias star formation diagnostics. Another possibility is that quasar radiation and / or quasar-driven outflows are in fact suppressing on-going star formation. Developing star formation diagnostics that can distinguish between these scenarios in hosts of luminous AGN remains an interesting challenge.

\section{Conclusions}
\label{sec:conclusions}

The correlation between radio luminosity and narrow line gas kinematics in radio-quiet quasar host galaxies \citep{mull13, zaka14} suggests that there may be a physical connection between the two. This correlation has renewed the debate over the origin of the radio emission in radio-quiet quasars. One hypothesis, proposed by \citet{cond13} and others, cites star formation in the host galaxy as the driving mechanism behind the observed radio emission from radio quiet quasars. The median radio luminosity of radio-quiet obscured quasars is $\nu L_{\nu}$[1.4 GHz]$=1\times 10^{40}$ erg s$^{-1}$ \citep{zaka14}, which would require 300 $M_{\odot}$ yr$^{-1}$ worth of star formation \citep{bell03}. While this is not impossible, it seems fairly high, so in this paper we ask whether the actual amount of star formation in the host galaxies is sufficient to produce the observed radio emission. 

We use archival samples of star-forming galaxies to revisit calibrations of different star formation indicators: radio emission (eq. \ref{eq_ir_radio}), single-band infrared fluxes (eq. \ref{eq_sym1}-\ref{eq_sym2}), and PAH[11.3\micron] luminosities (eq. \ref{eq_shi}-\ref{eq_pah}). As long as uniform measurement methods are used -- which is particularly important for PAH emission -- we find excellent agreement between published calibrations and archival data, with a spread in different correlations of $\sim 0.2$ dex. With these star-formation calibrations in hand, we measure star formation rates in the hosts of quasars of different types at $z<1$ using photometric and spectroscopic data from the \spi\ and \her\ telescopes. 

Using infrared colors of type 2 quasars, we demonstrate that the bolometric luminosities of the objects in our samples -- estimated to range from $10^{45}$ to $10^{47}$ erg s$^{-1}$, Figure \ref{pic_dist} -- are dominated ($>50$\%) by quasar emission; however, even in objects with 80\% quasar contribution to the bolometric budget, more than half of the 160\micron\ emission is due to star formation because the spectral energy distribution of starlight-heated dust peaks at much longer wavelengths than the quasar-heated dust which is concentrated on much smaller spatial scales. Thus we conclude that while our 160\micron\ measurements strictly speaking yield upper limits on star formation, the measured rates are close to the actual values. This is confirmed by spectral energy distribution modeling of a subsample of type 2 quasars drawn from the upper decade of our luminosity distribution \citep{wyle15}. 

Using 160\micron\ fluxes in $\sim 245$ obscured and unobscured quasars, we find a broad distribution of star formation rates in quasar hosts, from median values of 6$M_{\odot}$ yr$^{-1}$ (in hosts of blue type 1 quasars) to 18$M_{\odot}$ yr$^{-1}$ (in type 2 hosts). The difference in star formation rates in hosts of type 1 and type 2 quasars has been seen in other observations \citep{kim06, zaka08, hine09, chen15} and remains a challenge to the standard geometric unification model. Although statistics of molecular gas observations are still limited, existing data suggest similar availability of cold gas in hosts of type 1 and type 2 quasars \citep{krip12, vill13}, and the relatively low star formation rates of type 1 quasar hosts is in tension with the availability of cold gas in them, suggesting low star formation efficiency \citep{ho05}.

We then use mid-infrared spectra of $\sim 160$ quasars of different types to estimate the star formation rates of their hosts using spectroscopic diagnostics, especially PAH emission. We find a broad distribution of star formation rates, with a median of $\la 30$ $M_{\odot}$ yr$^{-1}$. We find that PAH ratios in quasars differ from those in star forming galaxies, in excellent quantitative agreement with the findings of \citet{diam10}, and we use PAH[11.3\micron] as our primary spectroscopic star formation indicator. Further evaluating [NeII]$\lambda$12.8\micron\ and [NeIII]$\lambda$14.3\micron\ lines as star formation indicator, we find both to be strongly affected and likely dominated by quasar photo-ionization. 

Regardless of the method used to estimate star formation in quasar hosts, we find that even the most generously computed star formation rates are insufficient to explain the observed radio emission, by about an order of magnitude. Depending on the measurement method, we measure the ratio of observed radio luminosity to that predicted due to star formation alone $L_{\rm radio,obs}/L_{\rm radio,SF}$ of 0.6$-$1.3 dex. Thus radio emission in radio-quiet quasars is unlikely to be dominated by star formation in quasar hosts and is likely associated with the quasars. Our results are in agreement with other evaluations of the star formation contribution to the radio emission of radio quiet quasars \citep{lal10, harr14}. 

\begin{figure}
\includegraphics[scale=0.8, clip=true, trim=0cm 10cm 10cm 0cm]{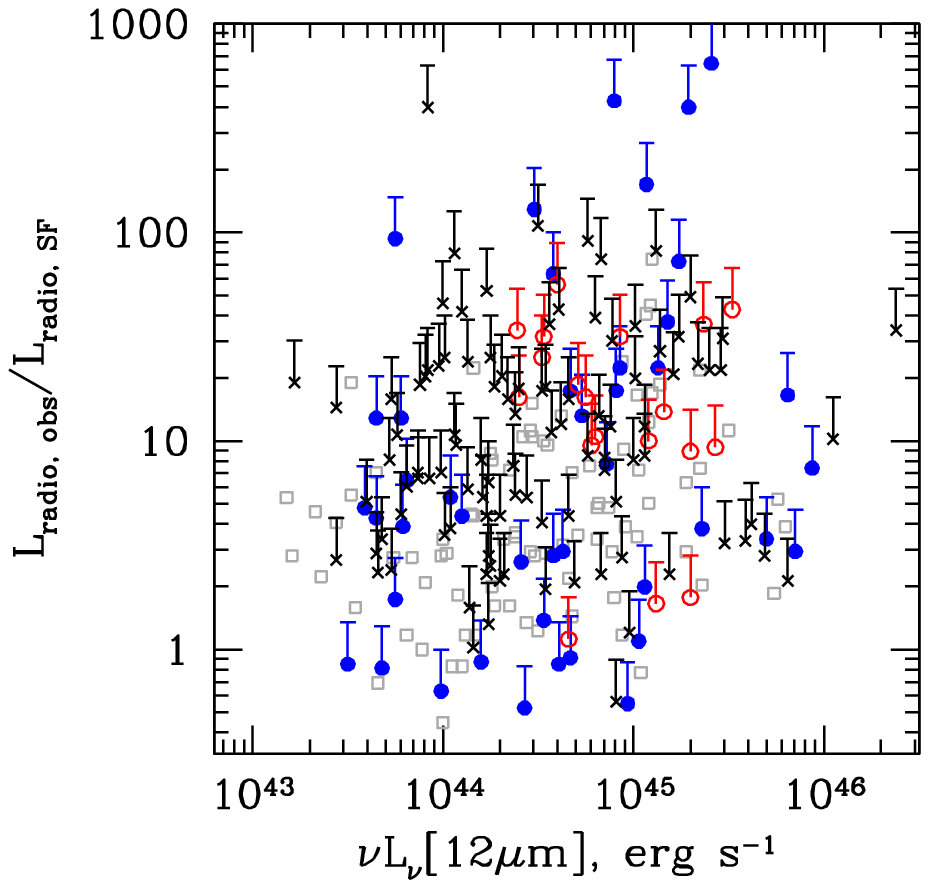}
\caption{The discrepancy between the observed radio emission and that expected from star formation as a function of the 12\micron\ luminosity. Grey points are for radio non-detections, black points for radio-detected type 2 quasars, blue for PG type 1s and red for 2MASS type 1s. The discrepancy between the predicted and the observed radio emission increases with infrared luminosity (Spearman rank probability of the null hypothesis of no correlation is P[NH]$=10^{-3}$). }
\label{pic_lum}
\end{figure}

\citet{rosa13} find that in AGN with $\nu L_{\nu}$[12\micron]$<10^{44}$ erg s$^{-1}$ both the radio emission and the 12\micron\ emission appear to closely follow the star-formation locus and they conclude that both 12\micron\ and radio emission are associated with star formation (see also \citealt{bonz15, pado15}). In Figure \ref{pic_lum} we show the discrepancy between the observed and the predicted radio emission as a function of the infrared luminosity. There is a slight trend of an increasing discrepancy toward higher 12\micron\ luminosity. Thus it is possible that for low-luminosity AGN ($\nu L_{\nu}$[12\micron]$\ll 10^{44}$ erg s$^{-1}$) the radio emission is a good measure of star formation rates in the host galaxy as \citet{rosa13} suggest, though some contribution to both radio and 12\micron\ emission from the AGN at $10^{43}<\nu L_{\nu}$[12\micron]$<10^{44}$ erg s$^{-1}$ appears likely in light of our results.

Regardless of the mechanism responsible for producing radio emission in the radio-quiet majority of quasars, this emission would be in addition to the radio emission produced by the host galaxy. This is well supported by Figures \ref{pic_photo}, \ref{pic_shi_pah} and \ref{pic_lum}, which show that rarely does the observed radio emission fall below the levels predicted to be due to star formation (and when it does, not by much). The objects whose $L_{\rm radio, obs}$ are closest to the $L_{\rm radio, SF}$ are those where the radio emission associated with the quasar is weaker than that due to star formation in the host galaxy (there are several such examples in this study and in \citealt{delm13}). Figure \ref{pic_lum} suggests that as the quasar luminosity increases, there are fewer such `radio-silent' objects. 

There are clear cases where quasar emission dominates all bands, including far-infrared \citep{hony11}, and more ambiguous objects in which the lowest temperature of the dust (as measured from the spectral energy distribution) is higher than typical values seen in star-forming galaxies \citep{sun14, tsai15} which could also be the case of a quasar dominating far-infrared emission. As we demonstrate in this paper, the radio emission in quasars is not a good estimate of star formation. Thus both $8-1000$\micron\ fluxes and radio emission -- used together or separately -- may strongly overestimate star formation rates of quasar hosts \citep{bart06, beel06}, and additional cross-checks or full spectral energy distribution decomposition are required \citep{wang08}. Calculating star formation rates in quasar hosts is further complicated by a possibly reduced star formation efficiency which would bias diagnostics based on cold gas detection \citep{ho05} and, for the unobscured star formation, by the strong contribution of direct or scattered light to the observed ultra-violet emission \citep{zaka05, obie15}.

While this work rules our star formation as the origin of most radio emission from quasars, we cannot distinguish between radio emission due to compact weak jets and radio emission due to wide-angle winds. Radiatively-driven winds can produce sufficient amount of radio emission only in quasars with $L_{\rm bol}\ga 3\times 10^{45}$ erg s$^{-1}$, which appears to be the threshold luminosity for driving galaxy-wide winds \citep{veil13a, zaka14}. Precise measurement of the radio luminosity function of these objects at faint radio luminosities is difficult because they are rare, so fairly high-redshift sources need to be observed with high sensitivity. Recent such work indicating a break in the radio luminosity function of quasars at fixed optical luminosity \citep{kimb11, cond13} suggests that different mechanisms are likely responsible for the emission of radio-quiet and radio-loud sources. 

This observation, and the radio vs gas kinematics correlation \citep{mull13, zaka14}, lead us to favor radiatively-driven winds as the ultimate origin of the radio emission \citep{stoc92, wang08a, jian10, fauc12b, zubo12, zaka14}. In this scenario, relativistic particles are accelerated on the shocks driven into the interstellar medium by the expanding wind. It may be testable by multi-wavelength observations \citep{nims15}, radio spectral measurements \citep{blun07} and high-resolution radio imaging. The problem of distinguishing radio emission from compact jets from radio emission as a bi-product of radiatively driven has proven especially difficult because the two mechanisms are similar in terms of energetics \citep{zaka14} and because a radiatively driven wind can inflate bubbles which mimic double-lobed radio morphologies, making morphology an unreliable jet / wind diagnostic \citep{harr15}. 

We have demonstrated that quasars and star-forming galaxies lie on {\it different far-infrared / radio correlations}: quasars show an order of magnitude more radio emission than do star-forming galaxies with the same 160\micron\ luminosities. However, multiple groups have demonstrated that quasars, star-forming galaxies and composite sources may lie on the {\it same total (8-1000\micron) infrared / radio correlations} \citep{craw96}, though possibly the spread \citep{mori10} or the normalization \citep{ivis10} of the correlation may change slightly in the quasar-dominated regime. If radio emission of radio-quiet quasars is dominated by jets, then the infrared/radio correlation can only be explained by a strong coupling between accretion processes and jet production \citep{falc95}, but finding jets on the same infrared / radio correlation as star-forming galaxies is quite surprising. If radio emission of radio-quiet quasars is dominated by radiatively driven winds, then in both quasars and star-forming galaxies the radio emission is produced by relativistic particles accelerated on shocks. What is surprising in this case is to find quasars and star-forming galaxies to be converting the same fraction of their bolometric power into shocks via completely different mechanisms \citep{nims15}. 

\acknowledgments

NLZ is grateful to E.Quataert, C.-A. Faucher-Gigu\`ere, J.Nims and the referee J. Mullaney for useful discussions. NLZ and RCH acknowledge support from the Herschel Science Center under JPL contracts No. 1475252 and No. 1471850, respectively. KL is supported by the Johns Hopkins University Dean's Undergraduate Research Award. RCH acknowledges support from an Alfred P. Sloan Research Fellowship and the Dartmouth Class of 1962 Faculty Fellowship. LCH acknowledges support by the Chinese Academy of Science through grant No. XDB09030102 (Emergence of Cosmological Structures) from the Strategic Priority Research Program and by the National Natural Science Foundation of China through grant No. 11473002. MO is supported in part by World Premier International Research Center Initiative (WPI Initiative), MEXT, Japan, and Grant-in-Aid for Scientific Research from the JSPS (26800093). 

This research made use of Tiny Tim/Spitzer, developed by John Krist for the Spitzer Science Center. The Center is managed by the California Institute of Technology under a contract with the National Aeronautics and Space Administration (NASA). This research has made use of the NASA / IPAC Infrared Science Archive and NASA/IPAC Extragalactic Database (NED), which are operated by the Jet Propulsion Laboratory, California Institute of Technology, under contract with NASA. 

\bibliographystyle{apj}
\bibliography{master}

\clearpage
\begin{deluxetable}{l|l|l|l|l|l|l|l|l|l}
\tabletypesize{\scriptsize}
\tablecaption{160\micron\ photometry and photometric constraints on star formation\label{tab:photo}}
\tablehead{
ID & subsample & $z$ & $L$[12\micron] & $F_{\nu}$[160\micron] & SFR & $L_{\rm radio,SF}$ & $L_{\rm radio,obs}$ & comment}	
\startdata
SDSS J004340.12-005150.2 & T2 &  0.5849 &   45.04 & -360.3 &  567 &   40.34 &  -40.23 & Spitzer, calibration                               \\ 
SDSS J005009.81-003900.6 & T2 &  0.7276 &   46.05 &  112.7 &  300 &   40.03 &   41.04 & Spitzer, ULIRG                                     \\ 
SDSS J005621.72+003235.8 & T2 &  0.4840 &   45.24 &  -72.9 &   95 &   39.47 &   40.97 & Spitzer, type 2 quasar (PI Strauss)                \\ 
SDSS J010523.62+011321.4 & T2 &  0.2049 &   43.89 & -230.8 &   55 &   39.20 &  -39.20 & Spitzer, calibration                               \\ 
SDSS J012341.47+004435.9 & T2 &  0.3990 &   44.80 &  -75.8 &   69 &   39.32 &   40.91 & Spitzer, type 2 quasar (PI Strauss)                \\ 

\enddata
\tablecomments{
Table presented in full in the electronic edition; a small portion given here for guidance regarding format and content.\\
`Subsample' describes whether the object belongs to the type 2 sample (T2) or red type 1 sample (T1red). \her\ data for blue type 1 quasars are published by \citet{petr15}, and \her\ data for type 2 quasars will be available in Petric et al. (in prep.)
$L$[12\micron] is given in units of $\log$($\nu L_{\nu}$[12\micron], erg s$^{-1}$) and is set to -100 for objects undetected in both WISE-12\micron\ and WISE-22\micron. \\
$F_{\nu}$[160\micron] measured in mJy are positive for detections (we use 25\% as an estimate of the absolute uncertainty) and negative for 5$\sigma$ upper limits. \\
SFR gives the upper limit on the star formation rate in $M_{\odot}$ yr$^{-1}$ calculated in Section \ref{sec:rem_photo}.\\
$L_{\rm radio,SF}$ is the upper limit on the radio emission due to star formation, given in units of $\log$($\nu L_{\nu}$[1.4GHz], erg s$^{-1}$).\\
$L_{\rm radio,obs}$ is the observed radio luminosity (or in the case of negative values, 5$\sigma$ upper limits) in units of $\log$($\nu L_{\nu}$[1.4GHz], erg s$^{-1}$).\\
Comment column describes whether the object was in the \spi\ or in the \her\ sample, whether it was targeted for pointed observations, and if so, why, or if it was covered serendipitously in observations of other targets (with the survey name given in parentheses if applicable). 
}
\end{deluxetable}

\clearpage
\begin{deluxetable}{l|l|l|l|l|l|l|l|l|l|l|l}
\tabletypesize{\scriptsize}
\tablecaption{IRS spectroscopy of type 2 quasars\label{tab:spec}}
\tablehead{
ID & $z$ & $S$[9.7\micron] & SL & $L_{\rm radio,obs}$ & $L$[NeII] & $L$[NeIII] & PAH & PAH & comment \\	
 &  &  & factor &  &  &  & [6.2\micron] & [11.3\micron] & }	
\startdata
SDSS J004252.56+153246.8 &  0.1175 &  0.64 & 1.000 &   39.56 &    1.19 &    1.84 &    6.09 &    7.20 & line-selected type 2 AGN \\ 
SDSS J005009.81-003900.6 &  0.7276 &  0.49 & 1.000 &   41.04 &  -10.48 &   21.14 &  -11.89 &  -32.60 & ULIRG \\ 
SDSS J005621.72+003235.8 &  0.4840 &  2.52 & 1.000 &   40.97 &    5.01 &     NaN &   -7.82 &   15.94 & type 2 quasar (PI Zakamska) \\ 
SDSS J012341.47+004435.9 &  0.3990 & -0.02 & 1.000 &   40.91 &    1.85 &     NaN &   -8.84 &   -5.15 & type 2 quasar (PI Zakamska) \\ 
SDSS J080224.35+464300.6 &  0.1206 &  0.36 & 1.300 &   39.69 &    0.80 &    1.55 &   -1.19 &    1.88 & line-selected type 2 AGN \\ 

\enddata
\tablecomments{
Table presented in full in the electronic edition; a small portion given here for guidance regarding format and content.\\
$S$[9.7\micron] is the dimensionless strength of the silicate feature (positive for absorption, negative for emission), similar to optical depth and defined in Section \ref{sec:spec}; typical systematic uncertainty in this measurement is 0.2 \citep{zaka10}.\\
`SL factor' is the multiplicative factor applied to the short-low IRS orders to bring them in agreement with the long-low orders. \\
$L$ of Ne and PAH emission features is given in units of $10^{42}$ erg s$^{-1}$, positive for detections, negative for 3$\sigma$ upper limits, `NaN' for lacking spectral coverage.\\
$L_{\rm radio,obs}$ is the observed radio luminosity (or in the case of negative values, 5$\sigma$ upper limits) in units of $\log$($\nu L_{\nu}$[1.4GHz], erg s$^{-1}$).\\
Comment column describes why the object was targeted for IRS observations. 
}
\end{deluxetable}

\end{document}